\documentclass[aps,prd,singlecolumn,11pt]{revtex4-1}

\usepackage{amsfonts,amssymb,amsmath}
\usepackage{graphicx}
\usepackage{braket}
\usepackage{mathtools}
\usepackage{titlesec}

\newcommand{\fsl}[1]{\ensuremath{\mathrlap{\!\not{\phantom{#1}}}#1}}

\begin{document}
	
	\title{Probing nonstandard lepton number violating\\interactions in neutrino oscillations}
	\author{Patrick D. Bolton$ ^{1,} $}
	\email{patrick.bolton.17@ucl.ac.uk}
	\author{Frank F. Deppisch$ ^{1,2,}$}
	\email{f.deppisch@ucl.ac.uk}
	
	\affiliation{$ ^{1} $Department of Physics and Astronomy, University College London, \\ \hbox{Gower Street, London WC1E 6BT, United Kingdom} \\
		$ ^{2} $\hbox{Institut f\"{u}r Hochenergiephysik, \"{O}sterreichische Akademie der Wissenschaften,} Nikolsdorfer Gasse 18, 1050 Wien, Austria}

\begin{abstract}
	\begin{center}
		\large{Abstract}
	\end{center}
	We discuss lepton number violating processes in the context of long-baseline neutrino oscillations. We summarise and compare neutrino flavour oscillations in quantum mechanics and quantum field theory, both for standard oscillations and for those that violate lepton number. When the active neutrinos are Majorana in nature, the required helicity reversal gives a strong suppression by the neutrino mass over the energy, $(m_{\nu}/E_{\nu})^{2}$. Instead, the presence of non-standard lepton number violating interactions incorporating right-handed lepton currents at production or detection alleviate the mass suppression while also factorising the oscillation probability from the total rate. Such interactions arise from dimension-six operators in the low energy effective field theory of the Standard Model. We derive general and simplified expressions for the lepton number violating oscillation probabilities and use limits from MINOS and KamLAND to place bounds on the interaction strength in interplay with the unknown Majorana phases in neutrino mixing. We compare the bounds with those from neutrinoless double beta decay and other microscopic lepton number violating processes and outline the requirements for future short- and long-baseline neutrino oscillation experiments to improve on the existing bounds.
\end{abstract}

\maketitle

\vspace{-0.5em}
\section{INTRODUCTION}
\vspace{-0.5em}

The seminal confirmation of neutrino flavour oscillations by the Super-Kamiokande and SNO experiments in 1998 and 2001, respectively, initiated a golden era in the experimental and theoretical studies of massive neutrinos in the Standard Model (SM) \cite{Fukuda:1998mi,Ahmad:2001an}. Five of the six (or eight) independent mixing parameters describing the three Dirac (or Majorana) neutrinos have now been pinned down by solar, atmospheric, accelerator and reactor neutrino oscillation experiments \cite{deSalas:2017kay}. 

There are still a few pieces of the neutrino mass puzzle that remain unknown. First, the value of the Dirac phase $ \delta $ which controls the magnitude of CP-violation in the neutrino sector. Second, the sign of the atmospheric squared mass difference $ \Delta m_{23}^{2} $ which will decipher the normal ordering (NO) or inverted ordering (IO) of the neutrinos. Both quantities will be determined in the coming years by accelerator and reactor oscillation experiments such as NO$\nu$A, DUNE, Hyper-Kamiokande and JUNO \cite{NOvA:2018gge, Acciarri:2015uup, Abe:2015zbg,An:2015jdp}. Neutrino oscillations are insensitive to the absolute scale of the neutrino masses $m_0$ -- beta decay experiments such as KATRIN aim to combine their results with precision measurements of the cosmic microwave background and large-scale galaxy clustering to place a stringent upper bound on this scale \cite{Osipowicz:2001sq, Capozzi:2017ipn, Lesgourgues:2012uu, Vagnozzi:2017ovm}.

Last but by no means the least is the question of the fundamental nature of the light neutrinos. For massive neutrinos -- or generally any massive fermion that is a singlet under gauge transformations -- it is possible to write two Lorentz invariant mass terms. The first, proposed by Dirac, is the same as that of the charged fermions. The second, proposed by Majorana, violates lepton number by two units \cite{Pal:2010ih,Akhmedov:2014kxa}. On the grounds of naturalness the latter has been favoured for some time, with effective models using the high scale of new physics (NP) to generate the tiny neutrino masses -- also known as the seesaw mechanism. Common UV-complete versions of this mechanism incorporate heavy Majorana right-handed (RH) neutrinos into the larger gauge structure of a grand unified theory \cite{GellMann:1980vs, Schechter:1980gr, Cheng:1980qt, Avignone:2007fu}. 

While these theories generically predict lepton number violating (LNV) phenomena, none have yet been observed in nature. If the light neutrinos are indeed Majorana, a standard explanation of this non-observation is to promote the non-anomalous combination of baryon number $ (B) $ minus lepton number ($L$) to a gauge group $ U(1)_{B-L} $. This symmetry is spontaneously broken at low energies and thereby suppresses ($ B-L $)-violating processes by the high symmetry-breaking scale \cite{Mohapatra:1980qe}. Other models which predict additional particles such as the Majoron and supersymmetric partners have been excluded at high significance by astronomical and collider observations \cite{Farzan:2002wx,Giudice:1992jg,Aad:2014wea, Aad:2015eda,Deppisch:2015qwa}. If the neutrinos are instead purely Dirac, the absence of an LNV signal could be explained if $B - L$ remains a global symmetry after the spontaneous symmetry breaking of a left-right (LR) symmetric model \cite{Bolton:2019bou}. There is also the intriguing possibility that neutrinos are quasi-Dirac \cite{Anamiati:2017rxw,Anamiati:2016uxp,Das:2017hmg}. 

The most promising current effort to detect an LNV process is via searches for neutrinoless double beta (0$ \nu\beta\beta $) decay which is the nuclear decay process $ (Z,A)\rightarrow (Z+2,A)+2e^{-}$. 
If observed, the absence of outgoing neutrinos implies there to be an internal neutrino propagator, which is possible only if the neutrino is its own antiparticle, i.e. if neutrinos are Majorana fermions. Even if unrelated new physics (NP) were to trigger $ 0\nu\beta\beta $ decay, a positive signal would automatically imply Majorana neutrinos via the Schechter-Valle black-box theorem \cite{Schechter:1981bd}. Current and future 0$ \nu\beta\beta $ experiments such as EXO-200, KamLAND-Zen, CUORE, GERDA-II and SuperNEMO are aiming to push the lower limit on the $0\nu\beta\beta$ decay half-life $ T^{0\nu\beta\beta}_{1/2} $ up to $10^{26}$ years for various nuclear isotopes  \cite{DellOro:2016tmg}. This lower limit can be converted to an upper limit on the neutrino mass parameter $ m_{\beta\beta} $ if the nuclear matrix element and associated systematic uncertainty are known for the relevant isotope. A large portion of the region in the $ m_{\beta\beta}-m_{0} $ parameter space corresponding to the quasi-degenerate arrangement of neutrino masses ($ m_{1}\simeq m_{2} \simeq m_{3} $) has been excluded at 90\% C.L. by the KamLAND-Zen and GERDA-II experiments. Future experiments will be able to exclude the region corresponding to the IO of neutrino masses ($ m_{3}\ll m_{1}<m_{2} $). The NO region ($ m_{1}<m_{2} \ll m_3$) can be probed unless the Majorana CP-violating (CPV) phases conspire to cancel the different contributions to $ m_{\beta\beta} $ \cite{Patrignani:2016xqp}.

Contributions to $ 0\nu\beta\beta $ decay from LNV physics beyond the SM have been studied extensively \cite{Mohapatra:1986su,Muto:1989cd,Babu:1995vh,Hirsch:1996qw,Pas:1997fx,Pas:2000vn,Prezeau:2003xn,Hirsch:2006yk,Duerr:2011zd,Deppisch:2012nb,Vergados:2012xy,Helo:2013dla,Helo:2013ika,Dev:2013vxa,Huang:2013kma,Deppisch:2017ecm,Simkovic2017,Graf:2018ozy,Cepedello:2018zvr}. Much of this past work has used an effective coefficient $ \varepsilon$ to parametrise the strength of the LNV non-standard interaction (NSI) with respect to the SM dimension-six Fermi interaction \cite{Miranda:2015dra}. Alternatively one can introduce Wilson coefficients for the relevant dimension-$ d $ combinations of SM fields, suppressed by the scale of NP raised to the power $ (4-d) $. A general analysis of LNV channels has been conducted in the effective field theory formalism for operators up to dimension eleven \cite{deGouvea:2007qla}. Experimental signatures of alternative LNV processes have been explored in the literature, for instance the decays $ K^{+}\rightarrow\pi^{-}\mu^{+}\mu^{+}$, $ \tau^{-}\rightarrow \pi^{-}\pi^{-}\mu^{+} $ and the lepton flavour and number violating (LNFV) muon conversion process $ \mu^{-}+(Z,A)\rightarrow e^+ +(Z-2,A)$ \cite{Dib:2000ce,Atre:2005eb,Patrignani:2016xqp}. While these channels are less sensitive to LNV compared to $ 0\nu\beta\beta $ decay, searches have been conducted by a variety of experiments \cite{Aaij:2011ex,Moulson:2013oga,CERNNA48/2:2016tdo}. The contributions of lepton number conserving (LNC) but potentially lepton flavour violating (LFV) NSI to standard processes such as meson decays, neutrino oscillations and neutrino scattering have also been considered exhaustively in the literature \cite{Fornengo:2001pm,Berezhiani:2001rs,Davidson:2003ha,Barranco:2005yy,Blennow:2007pu,Kopp:2007ne,Antusch:2008tz,Biggio:2009nt,Biggio:2009kv,Kopp:2010qt,Friedland:2011za,Ohlsson:2012kf,Ohlsson:2013nna,Esmaili:2013fva,Agarwalla:2014bsa,Girardi:2014gna,Miranda:2015dra,Farzan:2015hkd,Forero:2016cmb,Lindner:2016wff,Ludl:2016ane,Salvado:2016uqu,Dutta:2017nht,Shoemaker:2017lzs,Rodejohann:2017vup,Farzan:2017xzy,Esteban:2018ppq}. For processes which are insensitive to lepton number -- for example, the outgoing neutrino is not detected in charged pion decays $ \pi^{\pm}\rightarrow\ell_{\alpha}^{\pm} \overset{(-)}{\nu}^{}_{\hspace{-0.3em}\alpha}$ -- precision probes of SM predictions such as lepton universality and the unitarity of the Cabibbo-Kobayashi-Maskawa (CKM) matrix can be used to constrain both LNC and LNV NSI \cite{Biggio:2009nt}.

Much of the literature has so far only considered the effect of LNC NSI on neutrino oscillations. This is for good reason -- neutrino oscillation experiments are typically only concerned with neutrino flavour at production and detection, inferring the process $ \nu_{\alpha}\rightarrow\nu_{\beta} $ (or $ \bar{\nu}_{\alpha}\rightarrow\bar{\nu}_{\beta} $) from the accompanying charged lepton $\ell_{\alpha}^{+}$ ($\ell_{\alpha}^{-}$) at production and $\ell_{\beta}^{-}$ ($\ell_{\beta}^{+}$) at detection. In many cases there is no detector at the neutrino source to identify the initial composition of flavours -- this and the associated energy distribution must be inferred from separate measurements and Monte Carlo simulations \cite{Ferrari:2007zzc,Cao:2011gb,Aliaga:2016oaz}. Often there is also no way to determine the charge of the outgoing lepton at the far detector and therefore to discern the incoming lepton as a neutrino or antineutrino. This probe of lepton number is not a priority for most oscillation experiments because \textquoteleft$ \nu_{\alpha}\rightarrow\bar{\nu}_{\beta} $' is heavily suppressed if the mechanism is the standard Majorana neutrino helicity reversal \cite{Bahcall:1978jn,Schechter:1980gk,Li:1981um,Langacker:1998pv}. Like any other helicity suppression this introduces a factor of $ \sim (m_{\nu}/E_{\nu})^{2} $ to the rate of the process, where $ m_{\nu} $ is the neutrino mass scale and $E_{\nu}$ is the neutrino energy. Hence the small neutrino masses ($m_{\nu}\sim 0.1 ~\text{eV} $) and typical large oscillation experiment neutrino energies ($E_{\nu}\sim 5 ~\text{MeV} - 2 ~\text{GeV}$) combine to suppress the magnitude of the  \textquoteleft$ \nu_{\alpha}\rightarrow\bar{\nu}_{\beta} $' and \textquoteleft$ \bar{\nu}_{\alpha}\rightarrow\nu_{\beta} $' processes by $\sim 10^{-21}-10^{-16}$. On the other hand, the highly suppressed amplitude of such a process can be exploited -- any positive signal of LNV such as an excess of \textquoteleft wrong'-signed charged leptons at the far detector would then strongly imply NP. Experiments that have been sensitive to the charge of outgoing leptons, such as the long-baseline (LBL) oscillation experiment MINOS and the LBL reactor/solar oscillation experiment KamLAND, can thus be used to constrain LNV NSI.

In this paper we begin Sec. II with a brief discussion and derivation of the $ \nu_{\alpha}\rightarrow\nu_{\beta} $ oscillation probability in quantum mechanics (QM) and quantum field theory (QFT). Using the latter we will study the \textquoteleft$ \nu_{\alpha}\rightarrow\bar{\nu}_{\beta} $' process for Majorana neutrinos, obtaining the expected $ \sim (m_{\nu}/E_{\nu})^{2} $ suppression of the total rate. We also show that the total rate cannot be factorised into an oscillation probability in a similar way to $ \nu_{\alpha}\rightarrow\nu_{\beta} $ oscillations. In Sec. III we consider the impact of an LNV NSI -- rather than the well-studied LNC NSI -- on neutrino oscillations. We will demonstrate that the total rate is no longer suppressed and factorises when the chirality of the production and detection vertices are opposite. We write down a general expression for the non-standard oscillation probability and a simplified expression in the two-neutrino (2$ \nu $) mixing approximation, specifically for the $\nu_{\mu}-\nu_{\tau}$ sector. This allows us in Sec. IV A to use a limit from the MINOS experiment on the \textquoteleft$ \nu_{
\mu}\rightarrow\bar{\nu}_{\mu} $' process to place bounds on the simplified $2\nu$ parameter space of this effective model. We generalise to the three-neutrino (3$\nu$) mixing scheme and re-evaluate constraints from MINOS  in Sec. IV B along with those from the KamLAND experiment in Sec. IV C. We compare these constraints to those from microscopic LNV processes such as $0\nu\beta\beta$ decay, $\mu^- - e^+$ conversion and radiative neutrino masses in Sec. IV D. To conclude, we summarise our results and briefly outline the potential for future oscillation experiments with similar sensitivity to improve on these bounds.

\vspace{-0.5em}
\section{SUMMARY OF NEUTRINO OSCILLATIONS IN QUANTUM MECHANICS AND QUANTUM FIELD THEORY}
\vspace{-0.5em}

Neutrino oscillations are derived straightforwardly in QM. Their origin lies in the mismatch between kinetic and charged-current (CC) interaction terms in the SM Lagrangian,
\begin{equation} \label{eq:SML}
\mathcal{L}_{\mathrm{SM}} \supset \bigg(\frac{1}{2}\bigg)\bar{\nu}_{i}(i\fsl{\partial}-m_{i})\nu_{i} - \frac{g}{\sqrt{2}}\Big(\bar{\nu}_{\alpha}\gamma^{\mu}P_{L}\ell_{\alpha}W_{\mu} + \mathrm{H.c.}\Big)~,
\end{equation}
where $P_{L}$ is the left-handed (LH) chirality projector and we have employed four-component spinor notation for the fields.
The kinetic term is diagonal in the basis of definite neutrino mass, labelled by the index $ i $, while the interaction term is diagonal in the basis in which the neutrino and associated charged lepton have the same flavour, labelled by the index $ \alpha $. 

The Lagrangian in Eq. (\ref{eq:SML}) is valid for both Dirac and Majorana neutrinos (up to the indicated factor of 1/2 in the Majorana kinetic term). The former and latter are defined by
\begin{equation}\label{DirvMaj}
\nu^{D}=\nu_{L}+\nu_{R}~,~~~\nu^{M}=\nu_{L}+(\nu_{L})^{c}~.
\end{equation}
Here $ \nu^{D} $ has a new RH component $ \nu_{R} $, whereas $ \nu^{M} $ has the charge conjugate of its LH component (up to an arbitrary intrinsic charge parity which is commonly set to unity). For Dirac neutrinos the SM CC interaction term shown in Eq. (1) creates negative helicity neutrinos $ \ket{\nu(q,-)} $ and annihilates positive helicity antineutrinos $ \ket{\bar\nu(q,+)} $. For Majorana neutrinos $ \ket{\bar\nu(q,+)} $ is equivalent to $ \ket{\nu(q,+)} $. The creation and annihilation of the other two degrees of freedom in the Dirac case, or the \textquoteleft wrong' helicity degree of freedom in the Majorana case, are suppressed by $\sim (m_{\nu}/E_{\nu}) $ at the amplitude level \cite{Langacker:1998pv}.

The following discussion is valid in both cases because the interaction Lagrangian remains the same. The fields in the flavour basis can be rotated to those in the mass basis using the Pontecorvo-Maki-Nakagawa-Sakata (PMNS) mixing matrix $U$,
\begin{equation} \label{eq:PMNS}
\nu_{\alpha}(x)=\sum_{i}U_{\alpha i}~\nu_{i}(x)~.
\end{equation}
While oscillation experiments produce neutrinos in a particular flavour eigenstate $ \ket{\nu_{\alpha}} $, for oscillations to take place this must be a coherent superposition of the mass eigenstates. The time evolution of each massive state is governed by the Schr\"{o}dinger equation, resulting in
\begin{equation} \label{eq:unitary}
\ket{\nu_{\alpha}(t)}=\sum\limits_{i}U^{*}_{\alpha i}~e^{-iE_{\mathbf{q}}t}\ket{\nu_{i}(t_{0})}=\sum\limits_{i}\sum\limits_{\beta}U^{*}_{\alpha i}~e^{-iE_{\mathbf{q}}t}~U_{\beta i}^{}\ket{\nu_{\beta}(t_{0})}~,
\end{equation}
where $ E_{\mathbf{q}} = \sqrt{|\mathbf{q}|^{2}+m_{i}^2}\approx |\mathbf{q}|+\frac{m_{i}^{2}}{2|\mathbf{q}|}  $ is the energy of each ultra-relativistic massive neutrino with mass $ m_{i} $ and shared three-momentum $ \mathbf{q} $. An oscillation probability can be derived by computing the square of the overlap between the time-evolved initial flavour state $ \ket{\nu_{\alpha}(T)} $ and an arbitrary final flavour state $ \ket{\nu_{\beta}} $,
\begin{equation} \label{eq:oscprob}\
\begin{split}
\begin{aligned}
P_{\nu_{\alpha}\rightarrow\nu_{\beta}}(L,E_{\mathbf{q}})&=\big|\braket{\nu_{\beta}|\nu_{\alpha}(T)}\big|^{2}=\bigg|\sum\limits_{i}U_{\alpha i}^{*}U_{\beta i}^{}~e^{iE_{\mathbf{q}}L}\bigg|^{2} \\
&=\sum\limits_{i}|U_{\alpha i}^{}|^{2}|U_{\beta i}^{}|^{2}+2 ~\mathrm{Re}\sum\limits_{i<j}U_{\alpha i}^{*}U_{\beta i}^{}U_{\alpha j}^{}U_{\beta j}^{*}~e^{-i\frac{\Delta m_{ij}^{2}}{2E_{\mathbf{q}}}L}~,
\end{aligned}
\end{split}
\end{equation}
where we have taken $T\simeq L $ and $L$ is the oscillation baseline.

This simple derivation was the first attempt to quantify oscillation phenomena in a quantum mechanical framework \cite{Bilenky:1976yj,Eliezer:1975ja,Fritzsch:1975rz,Bilenky:1976yj}. It was soon realised, however, that this description relies on numerous unphysical assumptions, namely the use of plane wave states. Assuming the propagating $ \ket{\nu_{i}} $ states to be plane waves of equal momenta $ \mathbf{q} $ forces the external particles at production to have definite energies and momenta.  Energy-momentum conservation at production is then in tension with the creation of three $\ket{\nu_i}$ with different energies $ E_{\mathbf{q}}=\sqrt{|\mathbf{q}|^{2}+m_{i}^{2}} $ \cite{Akhmedov:2009rb}. While it is possible to derive Eq. (\ref{eq:oscprob}) without the equal momentum assumption, an overall uncertainty in the energy and momentum of the neutrino mass eigenstates is still a necessary component \cite{Winter:1981kj,Giunti:1991ca}.

Any rigorous treatment of oscillation in QM must therefore describe the $ \ket{\nu_{i}} $ states with wave packets, which is also known as the internal wave packet model \cite{Nussinov:1976uw, Kayser:1981ye, Kiers:1995zj}. As in Eq. (\ref{eq:oscprob}), the oscillation probability is proportional to the overlap of the wave packets. The loss of coherence seen at long distance in oscillation experiments qualitatively arises from dispersion of the wave packets, which propagate at different group velocities $ \mathbf{v}=\frac{\partial E_{\mathbf{q}}}{\partial \mathbf{q}}\big|_{\mathbf{q}=\mathbf{Q}} $ with $ \mathbf{Q} $ the mean momentum \cite{Giunti:1997wq,Zralek:1998rp,Beuthe:2001rc,Akhmedov:2010ms}. 

Despite these improvements there are still fundamental issues with this picture. Firstly, it is not apparent what shape the neutrino wave packets should take. Secondly, a finite energy-momentum uncertainty requires the localisation in space and time of the production and detection interaction processes, which is not accounted for \cite{Beuthe:2001rc}. It is also difficult to define a meaningful Fock space for the flavour eigenstate neutrinos \cite{Giunti:2003dg}. Finally, the QM approach does not consider the possible entanglement between the outgoing $\nu_i$ and $\ell_{\alpha}^{\pm}$ at production \cite{Cohen:2008qb}.

\begin{figure}[t]
	\includegraphics[width=6cm]{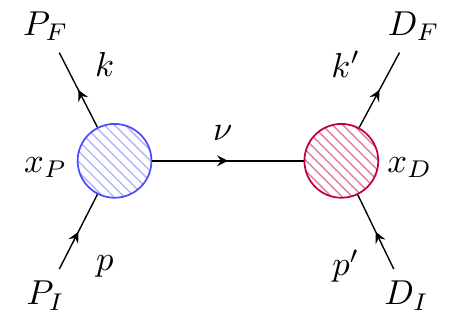}
	\includegraphics[width=6cm]{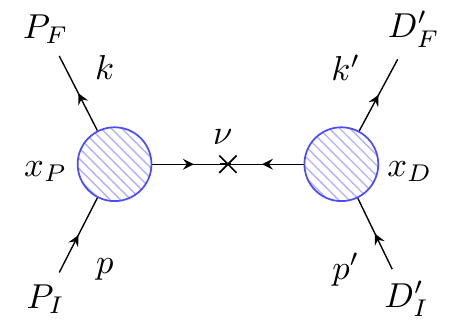}
	\caption{Generic Feynman diagrams depicting the $\nu_{\alpha}\rightarrow\nu_{\beta}$ process (left) and \textquoteleft$\nu_{\alpha}\rightarrow\bar{\nu}_{\beta}$' process (right), which requires a helicity reversal if the interaction vertices are SM-like.}
\label{fig:genericosc}
\end{figure}

A fully consistent framework in which to study neutrino oscillations is QFT, also known as the external wave packet model \cite{Giunti:1993se}. In this formalism the entire production, propagation and detection process can be described by a macroscopic Feynman diagram, as shown in Fig. \ref{fig:genericosc}. The external interacting particles are wave packets centred on the production and detection points $x_P$ and $x_D$, while the neutrinos are described by an internal propagator. The neutrino wave packets used in QM can be derived from the external particle wave packets in QFT; in QFT, however, the coherence conditions at each stage in the process are explicit \cite{Grimus:1998uh,Cardall:1999ze,Akhmedov:2010ms}.

To calculate an oscillation probability in QFT we must first compute an overall rate. To do this one constructs the $S$-matrix element at first order in the Fermi coupling $ G_{F}/\sqrt{2} = g^{2}/8m_{W}^{2}$,
\begin{equation} \label{eq:genericamp}
\begin{split}
\begin{aligned}
i\mathcal{A}_{\nu_{\alpha}\rightarrow \nu_{\beta}}(T,\mathbf{L})=\bra{P_{F}, D_{F}}\hat{T}\bigg\{\int d^{4}x_{1}\int d^{4}x_{2}~\mathcal{L}_{P}(x_{1})~\mathcal{L}_{D}(x_{2})\bigg\}\ket{P_{I}, D_{I}}~,
\end{aligned}
\end{split}
\end{equation}
where $ P_{I}$, $D_{I}$, $P_{F} $ and $ D_{F} $ are the initial and final state particles at production and detection, respectively, and $x_1$ and $x_2$ are space-time points in the vicinity of $x_P$ and $x_D$. This tree level process is depicted to the left of Fig. \ref{fig:genericosc}. The external asymptotic states $ \psi\in\{P_{I}, D_{I}, P_{F}, D_{F}\} $ are now described by the wave packets
\begin{equation} \label{eq:momentumspace}
\ket{\psi}=\int [dp]~f^{(\psi)}_{\mathbf{P}_{\psi}}(\mathbf{p}) \ket{\psi,\mathbf{p}}~,~~~[dp]=\frac{d^{3}p}{(2\pi)^{3}\sqrt{2E}}~,
\end{equation}
where $ f^{(\psi)}_{\mathbf{P}_{\psi}}(\mathbf{p}) $ is the momentum distribution function of an external particle $\psi$ with mean momentum $ \mathbf{P}_{\psi} $. To create and annihilate an internal neutrino the production and detection Lagrangian terms must take the general form
\begin{equation}
\begin{split}
\begin{aligned}
\mathcal{L}_{P}(x)= \sum\limits_{i}U_{\alpha i}^{*}~\bar{\nu}_{i}(x)~\widetilde{\mathcal{L}}_{P}(x),~~\mathcal{L}_{D}(x)= \sum\limits_{i}\widetilde{\mathcal{L}}_{D}(x)~U_{\beta i}^{}~\nu_{i}(x)~,
\end{aligned}
\end{split}
\end{equation}
where $ \widetilde{\mathcal{L}}_{P}(x) $ and $ \widetilde{\mathcal{L}}_{D}(x) $ are the \textquoteleft reduced' Lagrangian interaction terms for production and detection with the neutrino fields and PMNS matrix elements removed.

The total rate is proportional to the spin-averaged $S$-matrix element squared,
\begin{equation}
\Gamma^{\mathrm{tot}}_{\nu_{\alpha}\rightarrow\nu_{\beta}}\propto\braket{|\mathcal{A}_{\nu_{\alpha}\rightarrow\nu_{\beta}}(T,\mathbf{L})|^{2}}~,
\end{equation}
which can be expanded as
\begin{equation} \label{eq:ampsquared}
\begin{split}
\begin{aligned}
\braket{\big|\mathcal{A}_{\nu_{\alpha}\rightarrow\nu_{\beta}}\big|^{2}}&=\mathrm{Tr}~\Big|\sum\limits_{i}U_{\alpha i}^{*}U_{\beta i}^{}~\mathcal{A}_{i}\Big|^{2} \\
&=\sum\limits_{i}|U^{}_{\alpha i}|^{2}|U_{\beta i}^{}|^{2}~\mathrm{Tr}\big|\mathcal{A}^{}_{i}\big|^{2}+2~\mathrm{Re}\sum\limits_{i<j}U_{\alpha i}^{*}U_{\beta i}^{}U_{\alpha j}^{}U_{\beta j}^{*}~\mathrm{Tr}\big[\mathcal{A}_{j}^{\dagger}\mathcal{A}^{}_{i}\big]~,
\end{aligned}
\end{split}
\end{equation}
where $ \mathrm{Tr} $ denotes the Dirac trace. The $ \mathcal{A}_{i} $ factors are
\begin{equation} \label{eq:finalamp}
\begin{split}
\begin{aligned}
\mathcal{A}_{i}=\int\frac{d^{4}q}{(2\pi)^{4}}~\widetilde{\Phi}_{D}(q)~\frac{\fsl q+m_{i}}{q^{2}-m_{i}^{2}+i\epsilon}~\widetilde{\Phi}_{P}(q)~e^{-iq\cdot(x_{D}-x_{P})}~,
\end{aligned}
\end{split}
\end{equation}
where $ \widetilde{\Phi}_{P} $ and $\widetilde{\Phi}_{D}$ are integrals quantifying the overlap of external wave packets at production and detection respectively, explicitly written as
\begin{equation} \label{eq:overlapintegrals}
\begin{split}
\begin{aligned}
\widetilde{\Phi}_{P}(q)&=\int d^{4}x_{1}'~e^{iq\cdot x_{1}'}\int[dp^{}][dk^{}]~f^{(P_{I})}_{\mathbf{P}}(\mathbf{p})~f^{(P_{F})*}_{\mathbf{K}}(\mathbf{k}) ~e^{-i(p-k)\cdot x_{1}'} ~\widetilde{\mathcal{M}}_{P}~, \\
\widetilde{\Phi}_{D}(q)&=\int d^{4}x_{2}'~e^{-iq\cdot x_{2}'}\int[dp'][dk']~f^{(D_{I})}_{\mathbf{P}'}(\mathbf{p}')~f^{(D_{F})*}_{\mathbf{K}'}(\mathbf{k}') ~e^{-i(p'-k')\cdot x_{2}'} ~\widetilde{\mathcal{M}}_{D}~,
\end{aligned}
\end{split}
\end{equation}
where $x'_1=x_1-x_P$, $x'_2=x_2-x_D$ and $\widetilde{\mathcal{M}}_P$, $\widetilde{\mathcal{M}}_D$ are \textquoteleft reduced' matrix elements defined by
\begin{equation} \label{ap:coherentsquareterm}
\begin{split}
\begin{aligned}
\widetilde{\mathcal{M}}_P = \bra{P_F,\mathbf{k}}\widetilde{\mathcal{L}}_P(x_1)\ket{P_I,\mathbf{p}}~,~~~\widetilde{\mathcal{M}}_D = \bra{D_F,\mathbf{k}'}\widetilde{\mathcal{L}}_D(x_2)\ket{D_I,\mathbf{p}'}~.
\end{aligned}
\end{split}
\end{equation}

The trace appearing in the interference term of Eq. (\ref{eq:ampsquared}) is now
\begin{equation} \label{ap:coherentsquareterm}
\begin{split}
\begin{aligned}
\mathrm{Tr}\big[\mathcal{A}^{\dagger}_{j}\mathcal{A}_{i}]&=\int \frac{d^{4}q}{(2\pi)^{4}}\int\frac{d^{4}q'}{(2\pi)^{4}}\frac{\mathrm{Tr}\big[\underline{\widetilde{\Phi}}_{D}\widetilde{\Phi}_{D}\big(\fsl q+m_{i})\widetilde{\Phi}_{P}\underline{\widetilde{\Phi}}_{P}\big(\fsl{ q}'+m_{j})\big]}{(q^{2}-m_{i}^{2}+i\epsilon)(q'^{2}-m_{j}^{2}+i\epsilon)}~e^{-i(q-q')\cdot(x_{D}-x_{P})}~,
\end{aligned}
\end{split}
\end{equation}
where the underline of $ \underline{\widetilde{\Phi}}_{P,D} $ is shorthand for $ \widetilde{\Phi}_{P,D}^{\dagger}\gamma^{0} $. We now require the well-known limit of the $d^{3}q$ integrals as $ L\rightarrow\infty $, first given in Ref. \cite{Grimus:1996av}. If $ \psi(\mathbf{q}) $ is a twice differentiable function, for large $ L=|\mathbf{L}| $ and $ A_{i}> 0 $ the integral
\begin{equation} \label{eq:gs}
\int \frac{d^{3}q}{(2\pi)^{3}} ~\frac{\psi(\textbf{q})~e^{i\textbf{q}\cdot\textbf{L}}}{A_{i}-\textbf{q}^{2}+i\epsilon}=-\frac{1}{4\pi L}~\psi\Big(\sqrt{A_{i}}~\frac{\mathbf{L}}{L}\Big)~e^{i\sqrt{A_{i}}L} +\mathcal{O}\Big(\big(\sqrt{A_{i}}L\big)^{-\frac{3}{2}}\Big)~.
\end{equation}
For $ A_{i}<0 $ the integral falls as $ L^{-2}$ and can be neglected. Applying Eq. (\ref{eq:gs}) to Eq. (\ref{ap:coherentsquareterm}) with $ A_{i}=E_{\mathbf{q}}^{2}-m_{i}^{2} $ gives
\begin{equation} \label{eq:traceexp}
\begin{split}
\begin{aligned}
\mathrm{Tr}\big[\mathcal{A}^{\dagger}_{j}\mathcal{A}_{i}]=\frac{1}{64\pi^{4}L^{2}}\int dE_{\mathbf{q}}\int dE'_{\mathbf{q}}~\mathrm{Tr}\big[\underline{\widetilde{\Phi}}_{jD}\widetilde{\Phi}_{iD}(\fsl{q_{i}}+m_{i})\widetilde{\Phi}_{iP}\underline{\widetilde{\Phi}}_{jP}(\fsl{q_{j}}+m_{j})\big]e^{i(|\mathbf{q}_{i}|-|\mathbf{q}_{j}|)L}~,
\end{aligned}
\end{split}
\end{equation}
which has effectively set the neutrinos to be on-shell. As has been stated before in the literature \cite{Delepine:2009qg,Delepine:2009am}, if the production and detection processes are of the same chirality, the trace on the right-hand side of Eq. (\ref{eq:traceexp}) can be factorised in the relativistic limit ($ m_{i}\approx 0 $) as
\begin{equation} \label{eq:fact}
\begin{split}
\begin{aligned}
\mathrm{Tr}\big[\underline{\widetilde{\Phi}}_{jD}\widetilde{\Phi}_{iD}(\fsl{q_{i}}+m_{i})\widetilde{\Phi}_{iP}\underline{\widetilde{\Phi}}_{jP}(\fsl{q_{j}}+m_{j})\big]\approx\mathrm{Tr}\big[\widetilde{\Phi}_{iD}(\fsl{q_{i}}+m_{i})\underline{\widetilde{\Phi}}_{jD}\big]\mathrm{Tr}\big[\underline{\widetilde{\Phi}}_{jP}(\fsl{q_{j}}+m_{j})\widetilde{\Phi}_{iP}\big]~.
\end{aligned}
\end{split}
\end{equation}
Furthermore, writing $ (\fsl q+m) $ as a spinor sum and expanding $ |\mathbf{q}|=\sqrt{E_{\mathbf{q}}^{2}-m_{i}^{2}} $ for $ m_{i}\approx 0 $, i.e.
\begin{equation} \label{eq:sup1}
\begin{split}
\begin{aligned}
P_{L}(\fsl q+m_{i})P_{R}&=u_{iL}(q,-)\bar{u}_{iL}(q,-)+u_{iL}(q,+)\bar{u}_{iL}(q,+) \\
& = \big(q^{0}+|\mathbf{q}|\big)
\begin{psmallmatrix}
0 & 0 & 0 & 0 \\
0 & 0 & 0 & 1 \\
0 & 0 & 0 & 0 \\
0 & 0 & 0 & 0
\end{psmallmatrix} +
\big(q^{0}-|\mathbf{q}|\big)
\begin{psmallmatrix}
0 & 0 & 1 & 0 \\
0 & 0 & 0 & 0 \\
0 & 0 & 0 & 0 \\
0 & 0 & 0 & 0
\end{psmallmatrix}\\
& \approx 2E_{\mathbf{q}}\bigg\{1-\bigg(\frac{m_{i}}{2E_{\mathbf{q}}}\bigg)^{2}\bigg\}
\begin{psmallmatrix}
0 & 0 & 0 & 0 \\
0 & 0 & 0 & 1 \\
0 & 0 & 0 & 0 \\
0 & 0 & 0 & 0
\end{psmallmatrix} +
2E_{\mathbf{q}}\bigg(\frac{m_{i}}{2E_{\mathbf{q}}}\bigg)^{2}
\begin{psmallmatrix}
0 & 0 & 1 & 0 \\
0 & 0 & 0 & 0 \\
0 & 0 & 0 & 0 \\
0 & 0 & 0 & 0
\end{psmallmatrix}~,
\end{aligned}
\end{split}
\end{equation}
we can see that for $ \nu_{\alpha}\rightarrow\nu_{\beta} $ the propagation of positive helicity neutrinos is doubly suppressed by $ (m_{i}/2E_{\mathbf{q}})^{2} $ compared to the propagation of negative helicity neutrinos.

Neglecting the positive helicity neutrino contribution to the spinor sum in Eq. (\ref{eq:sup1}), the negative helicity spinors can be absorbed into the overlap integrals on either side in the traces of Eq. (\ref{eq:fact}):
\begin{equation}
\Phi_{P}\equiv\frac{\bar{u}^{(-)}_{jL}(q)}{\sqrt{2E_{\mathbf{q}}}}~\widetilde{\Phi}_{iP},~~\Phi_{P}^{*}\equiv\underline{\widetilde{\Phi}}_{jP}~\frac{u^{(-)}_{jL}(q)}{\sqrt{2E_{\mathbf{q}}}}~,
\end{equation}
where in the relativistic limit the mass eigenstate indices can be neglected \cite{Akhmedov:2010ms}. The trace appearing in the interference term can now be expressed as
\begin{equation}
\mathrm{Tr}\big[\mathcal{A}^{\dagger}_{j}\mathcal{A}_{i}]=\frac{1}{64\pi^{4}L^{2}}\int dE_{\mathbf{q}}\int dE'_{\mathbf{q}}~\braket{|\Phi_{P}|^{2}}\braket{|\Phi_{D}|^{2}}4E^{}_{\mathbf{q}}E'_{\mathbf{q}}~e^{i(|\mathbf{q}_{i}|-|\mathbf{q}_{j}|)L}~,
\end{equation}
and we see that the contributions from production, propagation and detection have factorised at the squared amplitude level. The total rate for the $ \nu_{\alpha}\rightarrow\nu_{\beta} $ process is now related to the differential production flux, oscillation probability and detection cross section as
\begin{equation} \label{eq:totalintermsofprodpropdet}
\Gamma_{\nu_{\alpha}\rightarrow\nu_{\beta}}^{\mathrm{tot}}(L,E_{\mathbf{q}})=\frac{1}{4\pi L^{2}}\int dE_{\mathbf{q}}~ \frac{d\Gamma_{\nu_{\alpha}}^{\mathrm{prod}}(E_{\mathbf{q}})}{ dE_{\mathbf{q}}}\cdot
P_{\nu_{\alpha}\rightarrow\nu_{\beta}}(L,E_{\mathbf{q}})\cdot \sigma^{\mathrm{det}}_{\nu_{\beta}}(E_{\mathbf{q}})~,
\end{equation}
where we have neglected experimental considerations such as the detector efficiency and fiducial volume \cite{Ankowski:2016jdd}. The oscillation probability can now be solved for through rearrangement of Eq. (\ref{eq:totalintermsofprodpropdet}). As Ref. \cite{Akhmedov:2010ms} shows, in the case of continuous fluxes of incoming particles, the differential production flux and detection cross section have the proportionality
\begin{equation} \label{eq:fluxcrosssec}
\frac{d\Gamma^{\mathrm{prod}}_{\nu_{\alpha}}}{dE_{\mathbf{q}}}\propto \sum\limits_{i}|U_{\alpha i}|^{2}~\braket{|\Phi_{P}|^{2}}~E_{\mathbf{q}} ~|\mathbf{q}_{i}|~,~~~ \sigma^{\mathrm{det}}_{\nu_{\beta}}\propto\sum\limits_{j}|U_{\beta j}|^{2}~\braket{|\Phi_{D}|^{2}}~E_{\mathbf{q}} ~|\mathbf{q}_{j}|^{-1}~.
\end{equation}
Using that $\braket{|\Phi_{P}|^{2}}$ and $\braket{|\Phi_{D}|^{2}}$ are independent of the mass $ m_{i} $, while also taking $ |\mathbf{q}_{i}|\approx|\mathbf{q}_{j}| $ in the relativistic or quasi-degenerate mass limit (equivalent to the inequality $ \big||\mathbf{q}_{i}|-|\mathbf{q}_{j}|\big|\ll |\mathbf{q}_{i}|,|\mathbf{q}_{j}| $), any dependence $ P_{\nu_{\alpha}\rightarrow\nu_{\beta}} $ has on the specific form of the wave packets cancels \cite{Akhmedov:2010ms}. In this limit $ P_{\nu_{\alpha}\rightarrow\nu_{\beta}}$ is given by Eq. (\ref{eq:oscprob}), confirming the result of the naive QM approach under the above conditions. 

For Majorana neutrinos the process \textquoteleft$ \nu_{\alpha}\rightarrow\bar{\nu}_{\beta} $' is possible and suppressed by $ (m_{i}/2E_{\mathbf{q}})^{2} $. To see this one can construct an amplitude like Eq. (\ref{eq:genericamp}) where the production and detection Lagrangian terms are the same
\begin{equation}
\mathcal{L}_{P}(x)=\mathcal{L}_{D}(x)=\sum\limits_{i}U_{\alpha i}^{*}~\bar{\nu}_{i}(x)~\widetilde{\mathcal{L}}(x)~,
\end{equation}
which is non-zero for Majorana neutrinos because $\bar{\nu}_i(x)$ both creates and annihilates $\ket{\nu_i}$. The Feynman diagram for this process is depicted in Fig. \ref{fig:genericosc} (right). The amplitude can again be squared and expressed as a sum of amplitudes $ \mathcal{A}_{i} $. Using the Majorana fermion Feynman rules of Ref. \cite{Denner:1992vza}, the amplitudes $ \mathcal{A}_{i} $ are identical to Eq. (\ref{eq:finalamp}) but with $ \Phi_{D}$ replaced by $ \Phi_{D}^{M}$. $ \Phi_{D} $ contains the reduced matrix element $ \widetilde{\mathcal{M}}_{D}\propto\bar{u}_{\beta}(p_{\beta})\Gamma $,  whereas $ \Phi_{D}^{M} $ contains $ \widetilde{\mathcal{M}}_{D}\propto\bar{u}_{\beta}(p_{\beta})C\Gamma C^{-1} $, where $ i\Gamma $ is the vertex factor for the detection process and $ C $ is the charge conjugation matrix. For left-handed SM CC interactions the trace $\mathrm{Tr}\big[\widetilde{\Phi}^{M}_{iD}(\fsl{q_{i}}+m_{i})\widetilde{\Phi}_{iP}\underline{\widetilde{\Phi}}_{jP}(\fsl{q_{j}}+m_{j})\underline{\widetilde{\Phi}}^{M}_{jD}\big]$
appearing in the squared amplitude now contains the factor
\begin{equation}
\begin{split}
\begin{aligned}
P_{R}(\fsl q_{i}+m_{i})P_{R}&=u_{iR}(q,-)\bar{u}_{iL}(q,-)+u_{iR}(q,+)\bar{u}_{iL}(q,+) \\
& = 2E_{\mathbf{q}}\bigg(\frac{m_{i}}{2E_{\mathbf{q}}}\bigg)
\begin{psmallmatrix}
0 & 0 & 0 & 0 \\
0 & 0 & 0 & 0 \\
0 & 0 & 0 & 0 \\
0 & 0 & 0 & 1
\end{psmallmatrix} +
2E_{\mathbf{q}}\bigg(\frac{m_{i}}{2E_{\mathbf{q}}}\bigg)
\begin{psmallmatrix}
0 & 0 & 0 & 0 \\
0 & 0 & 0 & 0 \\
0 & 0 & 1 & 0 \\
0 & 0 & 0 & 0
\end{psmallmatrix}
\end{aligned}
\end{split}
\end{equation}
squared, and so is singly suppressed by $ (m_{i}/2E_{\mathbf{q}})^{2} $ compared to the propagation of negative helicity neutrinos in Eq. (\ref{eq:sup1}) \cite{Xing:2013ty,Xing:2013woa,deGouvea:2002gf}.

We also see that in the relativistic limit the trace vanishes instead of factorising into components corresponding to production, oscillation, and detection. Strictly speaking it is therefore impossible to define an oscillation probability if helicity reversal is the dominant mechanism contributing to \textquoteleft$ \nu_{\alpha}\rightarrow\bar{\nu}_{\beta} $', only a probability for the entire process \cite{Delepine:2009qg}.

\vspace{-0.5em}
\section{NON-STANDARD LEPTON NUMBER VIOLATING INTERACTIONS}
\vspace{-0.5em}

We will now consider interactions at production and detection which are different from the usual SM CC interaction. CC-like and neutral current (NC)-like NSI which conserve lepton number, but introduce a new source of LFV to the SM have been studied extensively in the literature; see Refs. \cite{Miranda:2015dra,Farzan:2017xzy,Esteban:2018ppq} for recent reviews. Data from LBL experiments such as MINOS, NO$\nu$A, T2K and KamLAND and short-baseline (SBL) reactor experiments such as Daya Bay, RENO and Double Chooz have already been exploited to probe the flavour structure of the $\varepsilon$ coefficients parametrising the NSI \cite{Fornengo:2001pm,Friedland:2006pi,Blennow:2007pu,Altmannshofer:2018xyo,Aartsen:2017xtt,Adamson:2016yso,Liao:2017uzy}. Future prospects of next-generation experiments such as DUNE, T2HK and T2HKK have also been explored \cite{Oki:2010uc,deGouvea:2015ndi,Coloma:2015kiu,Masud:2015xva,Blennow:2016etl,Huitu:2016bmb,Fukasawa:2016lew,Masud:2016bvp,Liao:2016orc,Agarwalla:2016fkh,Bakhti:2016gic,Bischer:2018zcz}. It is solely the flavour of the charged leptons at detection that enables these constraints to be made, with the charge of the outgoing charged lepton being irrelevant. The Dirac or Majorana nature of the light neutrinos is therefore not probed.

\begin{table}[t]
	\begin{tabular}{l|l|ll}
		\hline \multicolumn{2}{l|}{~~~~~~~~~~~$\boldsymbol{\Delta L=0+\mathrm{H.c.}}$}& \multicolumn{2}{l}{~~~~~~~~~~~~~~$\boldsymbol{|\Delta L|=2+\mathrm{H.c.}}$}     \\ \hline \hline
		~$\mathcal{O}_{\nu edu}^{V,LL}$~ &  ~~$\left( \overline { \nu } _ { L p } \gamma ^ { \mu } e _ { L r } \right) \left( \overline { d } _ { L s } \gamma _ { \mu } u _ { L t } \right)$        & \multicolumn{1}{l|}{~$\mathcal { O } _ { \nu e d u } ^ { S , L L }$~~} &  ~~~~~$\left( \nu _ { L p } ^ { T } C e _ { L r } \right) \left( \overline { d } _ { R s } u _ { L t } \right)$\\
		~$\mathcal { O } _ { \nu e d u } ^ { V , L R }$~&    ~ $\left( \overline { \nu } _ { L p } \gamma ^ { \mu } e _ { L r } \right) \left( \overline { d } _ { R s } \gamma _ { \mu } u _ { R t } \right)$      & \multicolumn{1}{l|}{~$\mathcal { O } _ { \nu e d u } ^ { T , L L }$~~} & $\left( \nu _ { L p } ^ { T } C \sigma ^ { \mu \nu } e _ { L r } \right) \left( \overline { d } _ { R s } \sigma _ { \mu \nu } u _ { L t } \right)$  \\
		~$\mathcal { O } _ { \nu e d u } ^ { S , R R }$~&    ~~~~ $\left( \overline { \nu } _ { L p } e _ { R r } \right) \left( \overline { d } _ { L s } u _ { R t } \right)$      & \multicolumn{1}{l|}{~$\mathcal { O } _ { \nu e d u } ^ { S , L R }$~~} & ~~~~~$\left( \nu _ { L p } ^ { T } C e _ { L r } \right) \left( \overline { d } _ { L s } u _ { R t } \right)$ \\
		~$\mathcal { O } _ { \nu e d u } ^ { T , R R }$~&     $\left( \overline { \nu } _ { L p } \sigma ^ { \mu \nu } e _ { R r } \right) \left( \overline { d } _ { L s } \sigma _ { \mu \nu } u _ { R t } \right)$      & \multicolumn{1}{l|}{~$\mathcal { O } _ { \nu e d u } ^ { V , R L }$~~} & ~~$\left( \nu _ { L p } ^ { T } C \gamma ^ { \mu } e _ { R r } \right) \left( \overline { d } _ { L s } \gamma _ { \mu } u _ { L t } \right)$  \\
		~$\mathcal { O } _ { \nu e d u } ^ { S , R L }$~&   ~~~~~$\left( \overline { \nu } _ { L p } e _ { R r } \right) \left( \overline { d } _ { R s } u _ { L t } \right)$       & \multicolumn{1}{l|}{~$\mathcal { O } _ { \nu e d u } ^ { V , R R }$~~} & ~~$\left( \nu _ { L p } ^ { T } C \gamma ^ { \mu } e _ { R r } \right) \left( \overline { d } _ { R s } \gamma _ { \mu } u _ { R t } \right)$ \\
		\hline
	\end{tabular}
	\caption{LNC (left) and LNV (right) dimension-six CC-like interactions in the LEFT notation of Ref. \cite{Jenkins:2017jig}. The indices $p$, $r$, $s$ and $t$ denote flavour.}
\label{table:i}
\end{table}  

If an experiment was indeed sensitive to the sign of the charged lepton $\ell_{\beta}^{\pm}$ produced in the CC-like interaction of the incoming neutrino at detection, it would be able to distinguish between neutrinos and antineutrinos, or in other words negative and positive helicity Majorana neutrinos. This has been possible in the past -- a magnetised far detector was used to determine the charge from the curvature of tracks in the steel scintillator near and far detectors of MINOS. The prompt energy deposit from a $\ell_{\beta}^+$ and neutron capture on hydrogen, measurable for KamLAND, was also used as a distinct signal from the $\ell_{\beta}^-$ case \cite{Eguchi:2003gg}. It is therefore worth discussing the LNV equivalents of the CC-like and NC-like NSI discussed previously, as the non-observation of an excess of \textquoteleft wrong'-signed charged leptons by these experiments is able to say something about the possible size of an LNV equivalent $\varepsilon$ coefficient.

The operators to the left of Table \ref{table:i} trigger the standard LNC NSI studied in the literature, regardless of neutrinos being Dirac or Majorana. These are given in the notation of Refs. \cite{Jenkins:2017jig,Jenkins:2017dyc}, which enumerate the effective operators of dimension-$d\leq6$ generated below the electroweak scale. The authors determine the matching conditions and anomalous dimensions required to evolve the Wilson coefficients of these low energy effective field theory (LEFT) operators up to the usual SM effective operators satisfying $SU(3)_c \times SU(2)_L \times U(1)_Y$. The LNC operators in Table \ref{table:i} are matched to $\Delta L = 0 $ operators in the 59 operator basis of the SM effective field theory (SMEFT). The LNV operators to the right of Table \ref{table:i} can also trigger an interaction at detection for an incoming neutrino. It is not possible to match these operators to $\Delta L = 0 $ dimension-six SMEFT operators above the electroweak scale, but instead to dimension-seven $|\Delta L| = 2 $ operators with an additional Higgs field $H$ to conserve hypercharge $U(1)_Y$. These dimension-seven operators have been considered previously along with the complete list of LNV odd-dimensional effective operators up to dimension eleven in Ref. \cite{deGouvea:2007qla}. The operators in Table \ref{table:i} correspond to terms in $SU(2)_L$ expansions of the operators $\mathcal{O}_{3_a}$, $\mathcal{O}_{4_b}$ and $\mathcal{O}_{8}$. The dimension-seven operators have also been considered more specifically in Ref. \cite{Geng:2016auy}, but parametrised there in terms of a Wilson coefficient $\widetilde{C}^{ud}=C^{ud}/\Lambda_{\mathrm{NP}}^3$, where $\Lambda_{\mathrm{NP}}$ is the scale of NP.

In this paper we will adopt the following notation. A general Lagrangian, normalised to the Fermi interaction, and which can trigger an LNV interaction at production or detection, can be written as
\begin{equation}
\mathcal{L}_{P,D}=-\frac{G_{F}}{\sqrt{2}}~\bigg\{j^{}_{V-A}J^{\dagger}_{V-A}+\sum\limits_{\rho,\sigma}\varepsilon^{(\rho,\sigma)}~j^{}_{(\rho)} J_{(\sigma)}^{\dagger}\bigg\}+\mathrm{H.c.}~,
\end{equation}
where the second term includes all possible Lorentz contractions of the leptonic current  $ j_{(\rho)}=(\bar{\nu} \mathcal{O}^{(\rho)} \ell) $ and hadronic current $ J_{(\sigma)} = (\bar{u}\mathcal{O}^{(\sigma)}d) $, where $ (\rho,\sigma) $ run over the usual scalar ($ S $), pseudoscalar ($ P $), vector ($ V $), axial vector ($ A $) and tensor $ (T) $ Dirac structures. We choose linear combinations of these proportional to the chirality projection operators $ P_{L} $ and $ P_{R} $.

The $ \varepsilon^{(\rho,\sigma)} $ coefficients parametrise the strength of the NSI compared to $ G_{F}/\sqrt{2} $. We can define these coefficients as matrices in the flavour basis ($ \varepsilon^{(\rho,\sigma)} $) or mass basis  ($ \gamma^{(\rho,\sigma)} $) of the neutrino field, related by the rotation
\begin{equation} \label{eq:pmnstransform}
\varepsilon_{\beta\alpha}^{(\rho,\sigma)}\equiv\sum\limits_{i}U_{\alpha i}^{}\gamma_{\beta i}^{(\rho,\sigma)}~,
\end{equation}
where $U$ is the usual PMNS matrix. It is now straightforward to see the effect of these interactions on oscillations. If the production and detection leptonic currents are of the same chirality, $ \nu_{\alpha}\rightarrow\nu_{\beta} $ will be unsuppressed, while for Majorana neutrinos \textquoteleft$ \nu_{\alpha}\rightarrow\bar{\nu}_{\beta} $' will be suppressed by $ (m_{i}/2E_{\mathbf{q}})^{2} $. If the production and detection are of opposite chirality, \textquoteleft$ \nu_{\alpha}\rightarrow\bar{\nu}_{\beta} $' will now instead be factorisable and suppressed by $ |\varepsilon^{(\rho,\sigma)}|^{2} $. A commonly used example is a $ V-A $ (\textquoteleft$L$') leptonic current at production and a $ V+A $ (\textquoteleft$R$') leptonic current at detection arising from a LR symmetric model with an additional broken $ SU(2)_{R} $ gauge symmetry \cite{Mohapatra:1974hk}.

An LBL oscillation experiment sensitive to the sign of the outgoing lepton at the far detector, while not detecting the $\ell_{\alpha}^{\pm}$ at the production process (occurring in the beam pipe), could measure the ratio
\begin{equation} \label{eq:firstratio} R_{\alpha\beta}\equiv\frac{N_{\ell_{\beta}^{+}}}{N_{\ell_{\beta}^{-}}}=\frac{\Gamma_{\nu_{\alpha}\rightarrow\bar{\nu}_{\beta}}+\Gamma_{\bar{\nu}_{\alpha}\rightarrow\bar{\nu}_{\beta}}}{\Gamma_{\nu_{\alpha}\rightarrow\nu_{\beta}}+\Gamma_{\bar{\nu}_{\alpha}\rightarrow\nu_{\beta}}}~,
\end{equation}
where $N_{\ell_{\beta}^{\pm}}$ is the number of detected $\ell_{\beta}^{\pm}$ at the far detector. If we assume NP to be the cause of the first term in the numerator of Eq. (\ref{eq:firstratio}), and that this total rate is factorisable, the terms in $ R_{\alpha\beta} $ can be decomposed as
\begin{equation} \label{eq:indcomp}
\begin{split}
\begin{aligned}
R_{\alpha\beta}=\frac{\int dE_{\mathbf{q}}~\sum\limits_{\rho,\sigma}\bigg(\frac{d\Gamma_{\nu_{\alpha}}}{dE_{\mathbf{q}}}\cdot P^{(\rho,\sigma)}_{\nu_{\alpha}\rightarrow\bar{\nu}_{\beta}}\cdot\sigma_{\bar{\nu}_{\beta}}+\frac{d\Gamma_{\bar{\nu}_{\alpha}}}{dE_{\mathbf{q}}}\cdot P^{(\rho,\sigma)}_{\bar{\nu}_{\alpha}\rightarrow\bar{\nu}_{\beta}}\cdot\sigma^{}_{\bar{\nu}_{\beta}}\bigg)}{\int dE_{\mathbf{q}}~\sum\limits_{\rho,\sigma}\bigg(\frac{d\Gamma_{\nu_{\alpha}}}{dE_{\mathbf{q}}}\cdot P^{(\rho,\sigma)}_{\nu_{\alpha}\rightarrow\nu_{\beta}}\cdot\sigma^{}_{\nu_{\beta}}+\frac{d\Gamma_{\bar{\nu}_{\alpha}}}{dE_{\mathbf{q}}}\cdot P^{(\rho,\sigma)}_{\bar{\nu}_{\alpha}\rightarrow\nu_{\beta}}\cdot\sigma_{\nu_{\beta}}\bigg)}~,
\end{aligned}
\end{split}
\end{equation}
where the $ (\rho,\sigma) $ superscript denotes a NP effect. All probabilities are given this superscript because the process \textquoteleft$ \nu_{\alpha}\rightarrow\bar{\nu}_{\beta} $' reduces the number of neutrinos that undergo $ \nu_{\alpha}\rightarrow\nu_{\beta} $, and thus the sum $ \sum_{\beta}P^{(\rho,\sigma)}_{\nu_{\alpha}\rightarrow\nu_{\beta}} $ does not equal unity. The standard oscillation probability must be scaled by a normalisation factor depending on the NP and that ensures $ \sum_{\beta}(P^{(\rho,\sigma)}_{\nu_{\alpha}\rightarrow\nu_{\beta}}+P^{(\rho,\sigma)}_{\nu_{\alpha}\rightarrow\bar{\nu}_{\beta}})= 1 $,
\begin{equation} \label{eq:normandnotprobs}
\begin{split}
\begin{aligned}
P^{(\rho,\sigma)}_{\nu_{\alpha}\rightarrow\nu_{\beta}}= \frac{1}{\mathcal{N}_{\alpha}^{(\rho,\sigma)}}~ \bigg|\sum\limits_{i}U_{\alpha i}^{*}U_{\beta i}^{}~e^{-i\frac{m_{i}^{2} }{2E_{\mathbf{q}}}L}\bigg|^{2},~~
P_{\nu_{\alpha}\rightarrow\bar{\nu}_{\beta}}^{(\rho,\sigma)} = \frac{1}{\mathcal{N}_{\alpha}^{(\rho,\sigma)}}~\bigg|\sum\limits_{i}U_{\alpha i}^{*}~\gamma_{\beta i}^{(\rho,\sigma)}~e^{-i\frac{m_{i}^{2} }{2E_{\mathbf{q}}}L}\bigg|^{2}~,
\end{aligned}
\end{split}
\end{equation}
where the normalisation factor is
\begin{equation}
\begin{split}
\begin{aligned}
\mathcal{N}_{\alpha}^{(\rho,\sigma)}=
1+\sum\limits_{\beta}\bigg|\sum\limits_{i}U_{\alpha i}^{*}~\gamma_{\beta i}^{(\rho,\sigma)}~e^{-i\frac{m_{i}^{2} }{2E_{\mathbf{q}}}L}\bigg|^{2}~.
\end{aligned}
\end{split}
\end{equation}
Expanding the denominator in Eq. (\ref{eq:normandnotprobs}) it is straightforward to see that $ \sum_{\beta}P^{(\rho,\sigma)}_{\nu_{\alpha}\rightarrow\nu_{\beta}}\approx 1-\mathcal{O}(\varepsilon^{4})$ and $\sum_{\beta}P^{(\rho,\sigma)}_{\nu_{\alpha}\rightarrow\bar{\nu}_{\beta}}\approx \mathcal{O}(\varepsilon^{2})-\mathcal{O}(\varepsilon^{6}) $, so taking $ \varepsilon \rightarrow 0 $ recovers the SM prediction. We assume that the NP is a small effect, $ \varepsilon^{(\rho,\sigma)} \ll 1 $, and thus neglect this modification to the probabilities, i.e. set $ \mathcal{N}_{\alpha}^{(\rho,\sigma)}\approx 1 $. The factor $\mathcal{N}_{\alpha}^{(\rho,\sigma)}$ cancels in the ratio $R_{\alpha\beta}$ regardless. As another simplification we take the NSI coefficients to be real in this work.

\begin{figure}
	\includegraphics[width=7cm]{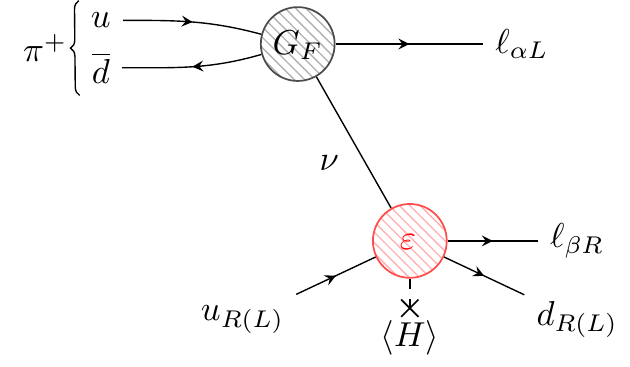}
	\includegraphics[width=7cm]{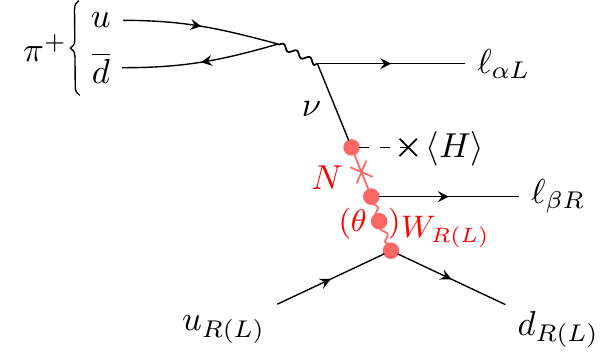}
	\caption{Left: simplified Feynman diagram of the oscillation process under consideration. Shown are the effective interactions at production and detection -- a low energy SM CC interaction and LNV NSI respectively. Right: a possible UV completion in a LR symmetric scenario.}
\label{fig:oscillationprocess}
\end{figure}

Expanding the probability to the right of Eq. (\ref{eq:normandnotprobs}) gives
\begin{equation} \label{eq:expand}
\begin{split}
\begin{aligned}
P^{(\rho,\sigma)}_{\nu_{\alpha}\rightarrow\bar{\nu}_{\beta}}&\approx\bigg|\sum\limits_{i}U_{\alpha i}^{*}~\gamma_{\beta i}^{(\rho,\sigma)}~e^{-i\frac{m_{i}^{2} }{2E_{\mathbf{q}}}L}\bigg|^{2}\\
&=\sum_{\lambda}F^{(\alpha)}_{\lambda}(L, E_{\mathbf{q}},\boldsymbol{\zeta},\boldsymbol{\eta})~\big(\varepsilon^{(\rho,\sigma)}_{\beta\lambda}\big)^{2} +\sum\limits_{\lambda<\lambda'} G^{(\alpha)}_{\lambda\lambda'}(L, E_{\mathbf{q}},\boldsymbol{\zeta},\boldsymbol{\eta}) ~\varepsilon_{\beta\lambda}^{(\rho,\sigma)}\varepsilon^{(\rho,\sigma)}_{\beta\lambda'}~,
\end{aligned}
\end{split}
\end{equation}
where we have rotated the $ \gamma^{(\rho,\sigma)}_{\beta i} $ back into the flavour basis using Eq. (\ref{eq:pmnstransform}) and $\lambda$, $\lambda'$ sum over flavour. The number of flavour and mass indices has been kept general and could include sterile states. The $ F^{(\alpha)}_{\lambda} $ and $ G^{(\alpha)}_{\lambda \lambda'} $ are functions of the baseline, neutrino energy, generic mixing parameters $ \boldsymbol{\zeta} $ and Majorana phases $\boldsymbol{\eta}$. For a general mixing matrix $U$ these functions take the form
\begin{equation} \label{eq:fandg}
\begin{split}
\begin{aligned}
&F^{(\alpha)}_{\lambda}(L,E_{\mathbf{q}},\boldsymbol{\zeta}) = \sum\limits_{i}|U_{\alpha i}|^2|U_{\lambda i}|^2+2~\mathrm{Re}\bigg\{\sum\limits_{i<j}U_{\alpha i}^*U_{\lambda i}^* U^{}_{\alpha j}U^{}_{\lambda j} e^{-i\frac{\Delta m_{ij}^2}{2E_{\mathbf{q}}}L}\bigg\}~,\\
&G^{(\alpha)}_{\lambda\lambda'}(L,E_{\mathbf{q}},\boldsymbol{\zeta})=2~\mathrm{Re}\bigg\{\sum_{i}|U_{\alpha i}|^2 U^{*}_{\lambda i} U^{}_{\lambda' i}\bigg\}\\
&~~~~~~~~~~~~~~~~~~~~~~+2~\mathrm{Re}\bigg\{\sum_{i<j} \big(U_{\alpha i}^*U_{\lambda i}^* U_{\alpha j} U_{\lambda' j}+U_{\alpha i}^*U_{\lambda' i}^* U_{\alpha j} U_{\lambda j}\big)e^{-i\frac{\Delta m_{ij}^2}{2E_{\mathbf{q}}}L}\bigg\}~.
\end{aligned}
\end{split}
\end{equation}
In the 3$\nu$ mixing scheme these are complicated functions of the three mixing angles, three squared splittings and two Majorana phases. If one were to consider atmospheric or accelerator \textquoteleft$ \nu_{\mu}\rightarrow\bar{\nu}_{\mu,\tau}$' oscillations, the 2$ \nu $ mixing approximation is valid because the oscillation Hamiltonian is dominated by the atmospheric mass splitting. The mixing matrix in this case is
\begin{equation}
U=\begin{pmatrix}
\cos\vartheta & \sin\vartheta~e^{i\eta}\\
-\sin\vartheta~e^{-i\eta} & \cos\vartheta
\end{pmatrix}~,
\end{equation} 
where $\vartheta$ is the single mixing angle (approximately corresponding to $\theta_{23}$) and $\eta$ is a Majorana phase \cite{Cabibbo:1963yz}, while the squared mass splitting is $\delta m^2$ (corresponding to $\Delta m_{23}^2$).
$F^{(\mu)}_{\lambda}$ and $G^{(\mu)}_{\lambda\lambda'}$ take on the simplified forms,
\begin{equation} \label{eq:functions}
\begin{split}
\begin{aligned}
F^{(\mu)}_{\mu}(L,E_{\mathbf{q}},\delta m^{2},\vartheta,\eta)&=1-\sin^{2}(2\vartheta) \sin^{2}\varphi~,\\
F^{(\mu)}_{\tau}(L,E_{\mathbf{q}},\delta m^{2},\vartheta,\eta)&=\sin^{2}(2\vartheta) \sin^{2}\varphi~,\\
G^{(\mu)}_{\mu\tau}(L,E_{\mathbf{q}},\delta m^{2},\vartheta,\eta)&=2\sin(2\vartheta)\sin^{2}\varphi~\big(\sin\eta \cot\varphi-\cos\eta\cos (2\vartheta)\big)~,
\end{aligned}
\end{split}
\end{equation}
where $ \varphi=\eta-\frac{\delta m^{2}L}{4E_{\mathbf{q}}} $. We will make use of these functions in the next section when studying the results from MINOS in the $2\nu$ mixing approximation.

\vspace{-0.5em}
\section{CONSTRAINTS ON LEPTON NUMBER VIOLATING NON-STANDARD INTERACTIONS}
\vspace{-0.5em}

We will now use the parametrisation discussed above to put constraints on the $\varepsilon$ coefficient parameter space of these LNV NSI. We will first discuss constraints from the MINOS experiment in the $2\nu$ mixing approximation in Sec. IV A, moving onto analysis of both MINOS and KamLAND in the $3\nu$ mixing scheme in Secs. IV B and C. In Sec. IV D we will compare these constraints to the more common limits from microscopic LNV processes such as $0\nu\beta\beta$ decay, $\mu^- - e^+$ conversion, rare meson decays and the radiative generation of neutrino masses. We summarise all limits in Table \ref{table:iii}.

\vspace{-0.5em}
\subsection{CONSTRAINTS FROM THE MINOS EXPERIMENT IN THE TWO NEUTRINO MIXING APPROXIMATION}
\vspace{-0.5em}

The MINOS experiment initially took data from 2005 to 2012 and used the low energy NuMI beam to detect neutrinos with a near detector at Fermilab and a far detector at a baseline of $L=735$ km from the source at the Soudan mine \cite{Pawloski:2016wwx}. After a break the experiment continued from 2013 to 2016 as MINOS+, using the medium energy NuMI beam \cite{Adamson:2013whj}. The experiment observed the disappearance of $ \nu_{\mu} $ produced from $ \pi^{+} $ decays (in the focusing beam configuration) and $ \bar{\nu}_{\mu} $ from $ \pi^{-} $ decays (defocusing), allowing the atmospheric mixing parameters dominating $ \nu_{\mu}\rightarrow\nu_{\mu} $ disappearance to be probed. The experiment also confirmed $ \nu_{e} $ and  $ \bar{\nu}_{e} $ appearance, constraining the reactor mixing angle $ \theta_{13} $. Most importantly for our discussion, charged lepton sign identification was possible in the near and far detectors through the use of 1.3 T toroidal magnetic fields -- $ \nu_{\mu} $, $ \bar{\nu}_{\mu} $, $ \nu_{e} $ and $ \bar{\nu}_{e} $ events could therefore be distinguished from the curvature of the outgoing $ \mu^{-} $, $ \mu^{+} $, $ e^{-} $ and $ e^{+} $ tracks, respectively.

Before MINOS began taking data, the expected fluxes of $\nu_{\mu}$ and $\bar{\nu}_{\mu}$ in the focusing and defocusing configurations were predicted to high precision from hadron production data and \textit{in situ} measurements. These were more recently updated in Ref. \cite{Aliaga:2016oaz}. In the focusing configuration the background of $\bar{\nu}_e$ produced from pion decays upstream of the target and avoiding deflection by the magnetic field is non-negligible and an important systematic error to correct  \cite{Kopp:2007cx}. There are also $\bar{\nu}_{e}$ produced downstream from secondary interactions in the beam pipe wall \cite{Danko:2009qw}.

The ratio in Eq. (\ref{eq:firstratio}) can be split into a signal part $ S_{\mu\mu} $ arising from the \textquoteleft$ \nu_{\mu}\rightarrow\bar{\nu}_{\mu} $' process and a background part $ B_{\mu\mu} $ arising from the standard oscillation of the background antineutrinos $ \bar\nu_{\mu}\rightarrow\bar{\nu}_{\mu} $. The MINOS analysis of Ref. \cite{Danko:2009qw} removes the predicted energy-dependent value of $ B_{\mu\mu} $ from the total measured $ R_{\mu\mu} $ and derives the constraint $ S_{\mu\mu}\lesssim 0.026 $. From
\begin{equation} \label{eq:signal}
S_{\mu\mu}\approx\frac{\int dE_{\mathbf{q}}~\sum\limits_{\rho,\sigma}\frac{d\Gamma_{\nu_{\mu}}}{dE_{\mathbf{q}}}\cdot P^{(\rho,\sigma)}_{\nu_{\mu}\rightarrow\bar{\nu}_{\mu}}\cdot\sigma^{}_{\bar{\nu}_{\mu}}}{\int dE_{\mathbf{q}}~\sum\limits_{\rho,\sigma}\frac{d\Gamma_{\nu_{\mu}}}{dE_{\mathbf{q}}}\cdot P^{(\rho, \sigma)}_{\nu_{\mu}\rightarrow\nu_{\mu}}\cdot\sigma^{}_{\nu_{\mu}}}\lesssim 0.026~,
\end{equation}
we can put corresponding constraints on the range of possible $ \varepsilon_{\mu\lambda}^{(\rho,\sigma)} $. The factorised form of Eq. (\ref{eq:signal}) of course assumes the chirality of the production and detection processes to be opposite. For example, we could consider a $ V-A $ leptonic current at production ($ \rho =  \sigma = L $) and a $ V+A $ leptonic current at detection ($\rho= R $). The two possible Lorentz contractions with the hadronic current are $ \sigma=R$ and $ \sigma=L $; in a LR symmetric model the former corresponds to the exchange of a $ W_{R} $ boson, the latter to $ W_{L}-W_{R} $ mixing if the boson mass eigenstates are mismatched, as depicted in Fig. \ref{fig:oscillationprocess}  \cite{Langacker:1998pv}. To simplify this work we will only consider these two cases, setting $ \varepsilon_{\mu\alpha}^{(\rho,\sigma)} $ for all other $ (\rho,\sigma) $ to zero and cleaning up the notation with $ \varepsilon_{\beta\alpha}^{(R,L)}= \varepsilon_{\beta\alpha}^{(R,R)}\equiv \varepsilon^{}_{\beta\alpha} $. It is worth noting that there is a subtle difference between the LNV NSI being at production and detection because the outgoing lepton $\ell_{\alpha}^{\pm}$ at production is not measured. Hence in theory one must sum over the different initial flavours as $\sum_{\alpha}S_{\alpha \mu} \lesssim 0.026$, and MINOS is sensitive to $\varepsilon_{e\lambda}$ and $\varepsilon_{\tau\lambda}$ . However, for the $\pi^{\pm}$ energies in the NuMI beam it is kinematically impossible to produce a $\tau^{\pm}$, and similar to standard $\pi^{\pm}$ decays the production of an electron is chirality suppressed to that of a muon. We therefore neglect this subtlety and assume that an NSI at production is probed in a similar way to that at detection.

\begin{figure}[t]
	{\hspace{-1em}\includegraphics[width=6.5cm]{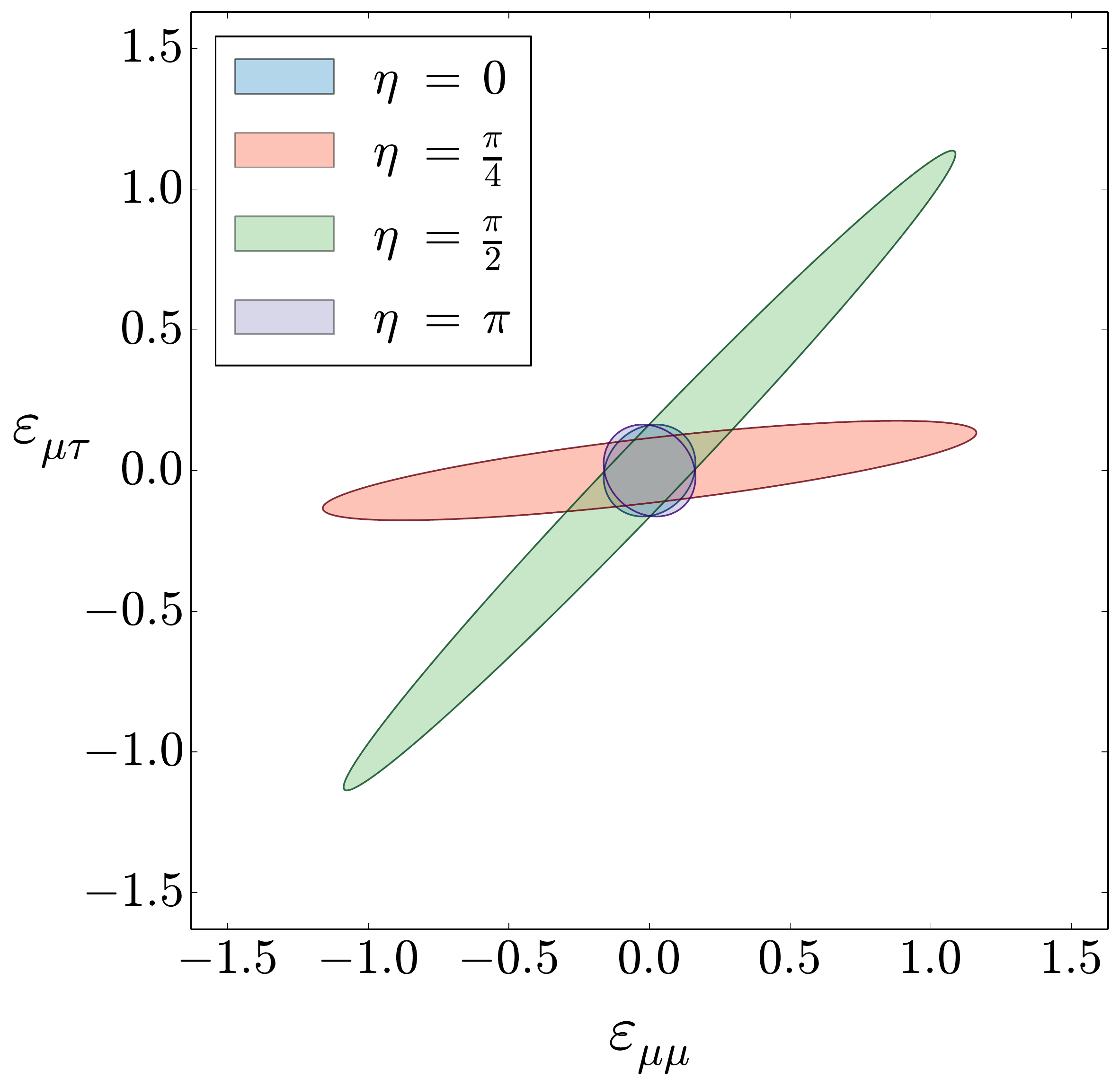}\hspace{0.5em}\includegraphics[width=6.65cm]{{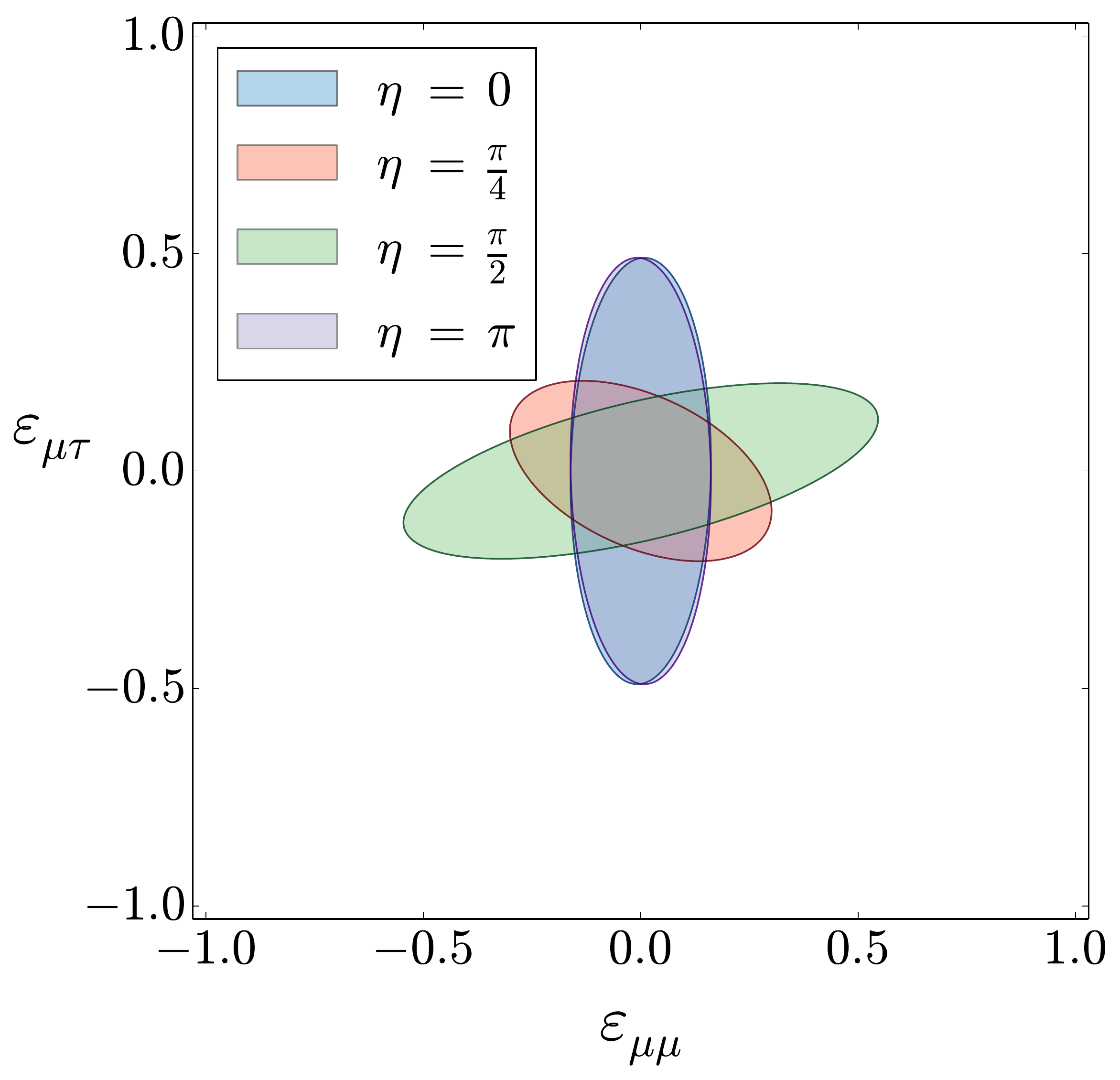}}}
	\caption{Allowed regions in the $ (\varepsilon_{\mu\mu},\varepsilon_{\mu\tau}) $ parameter space for fixed $ L/E_{\mathbf{q}} = 735/3 $ km GeV$^{-1}$ and four values of $ \eta $ in the 2$\nu$ mixing approximation of the $ \nu_{\mu}-\nu_{\tau} $ sector (left). Allowed region in $ \varepsilon_{\mu\mu}$ -- $\varepsilon_{\mu\tau} $ for fixed $ L$ = 735 km and again four values of $ \eta $, found by integrating over the NuMI beam neutrino energies (right).}
\label{fig:fixedversusintegrated}
\end{figure}

Working with the 2$ \nu $ mixing expansion of $ P_{\nu_{\mu}\rightarrow\bar\nu_{\mu}} $ given in Eqs. (\ref{eq:expand}) and (\ref{eq:functions}), Eq. (\ref{eq:signal}) can be used to put a constraint on the $ (\varepsilon_{\mu\mu},\varepsilon_{\mu\tau}) $ parameter space. To do this we integrate the NuMI $ \nu_{\mu} $ differential fluxes of Ref. \cite{Aliaga:2016oaz}, normalised probabilities and cross sections in the numerator and denominator of Eq. (\ref{eq:signal}) over 500 MeV bins in the range $0 - 20$ GeV. The cross sections
\begin{equation} \label{eq:csapprox}
\sigma_{\bar{\nu}_{\mu}p\rightarrow\ell_{\beta}^{+}n}(E_{\mathbf{q}})\backsimeq \sigma_{\nu_{\mu}n\rightarrow\ell_{\beta}^{-}p}(E_{\mathbf{q}})\backsimeq\frac{G_{F}^{2}|V_{ud}|^{2}}{\pi}\big(g_{\mathrm{V}}^{2}+3g_{\mathrm{A}}^{2}\big)E_{\mathbf{q}}^{2}~,
\end{equation}
are assumed to be equal in the quasi-elastic scattering limit \cite{Formaggio:2013kya}.

In Fig. \ref{fig:fixedversusintegrated} (left) we first plot the allowed regions in the $ (\varepsilon_{\mu\mu},\varepsilon_{\mu\tau}) $ parameter space for fixed $ L/E_{\mathbf{q}}=735/3 $ km GeV$ ^{-1} $, using best fit values for $  \delta m^2\approx\Delta m_{23}^{2}$, $ \vartheta\approx\theta_{23}$ (in the NO scheme), $ G_{F} $, $ V_{ud} $, $ g_{V} $, $ g_{A} $ and four different values of the Majorana phase $ \eta $  \cite{Patrignani:2016xqp}. This is equivalent to assuming the $ \nu_{\mu} $ flux to be sharply peaked at 3 GeV and evaluating the oscillation probability and cross section at this energy. Appreciable constraints are possible for $ \eta = 0 $ and $ \pi $, and are marginally better in the $ \varepsilon_{\mu\tau} $ direction. They are of order $ |\varepsilon_{\mu\mu}|\lesssim 0.2 $ and $ |\varepsilon_{\mu\tau}|\lesssim 0.1 $. When $ \eta = (n+1/2)\pi $ for $ n\in \mathbb{Z} $ a specific direction in the parameter space alleviates the constraints. The narrow ellipse arises because $ F^{(\mu)}_{\tau}\ll 1 $ for the best fit parameters and these particular values of $ \eta $. In Fig. \ref{fig:fixedversusintegrated} (right) we depict the allowed regions after the full numerical integration of the numerator and denominator of $S_{\mu\mu}$. For $\eta=\pi/4$ and $\pi/2 $ the narrow bands of allowed values are reduced to ellipses more similar to the ellipses at $ \eta =0$ and $\pi$. The orientations of the ellipses also change marginally. Upper bounds are in the ranges $|\varepsilon_{\mu\mu}|\lesssim 0.2-0.5$ and $|\varepsilon_{\mu\mu}|\lesssim 0.2-0.6$. 
\begin{figure}[t]
	{\includegraphics[width=7cm]{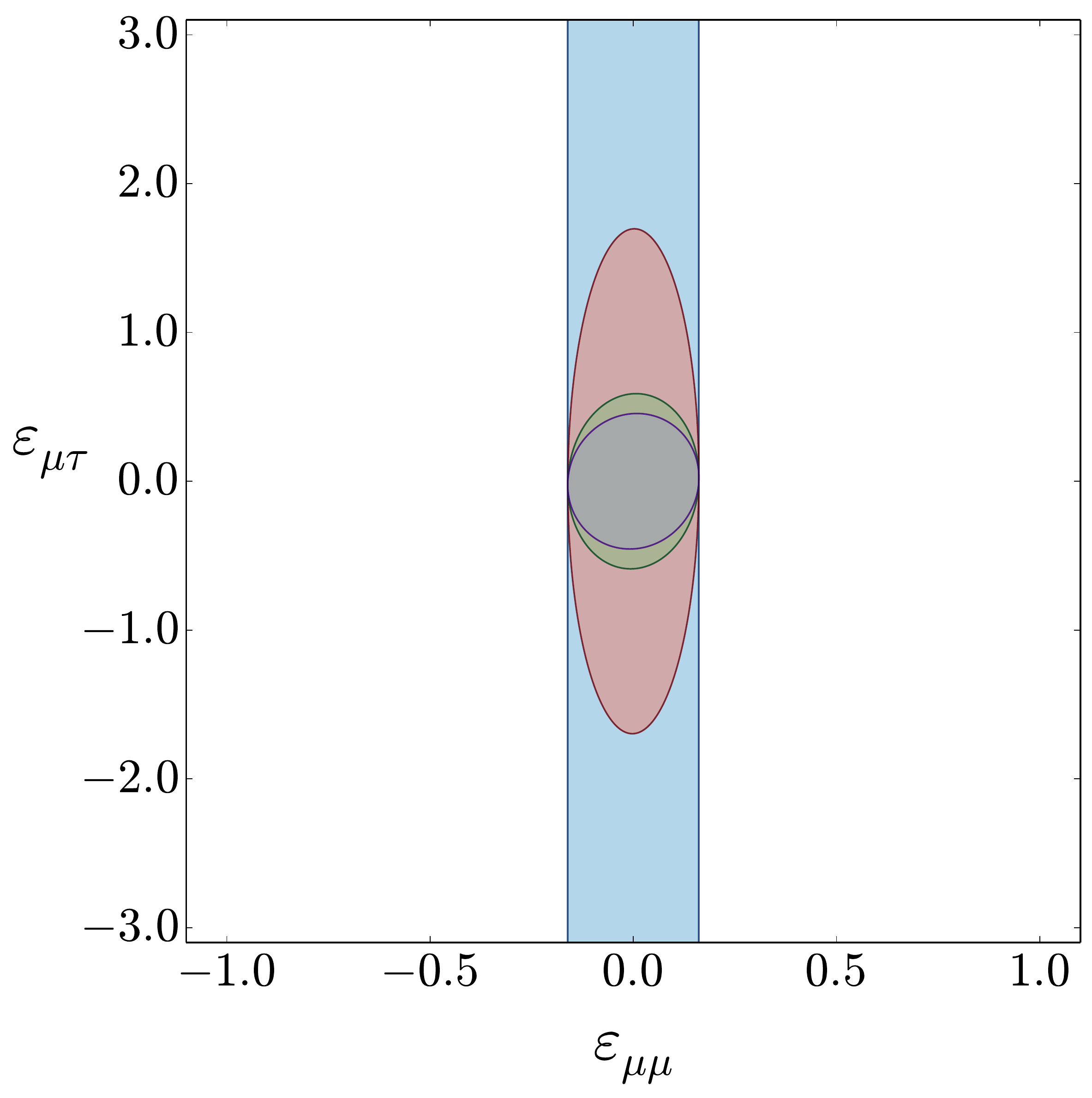}\hspace{0.5em}\includegraphics[width=7cm]{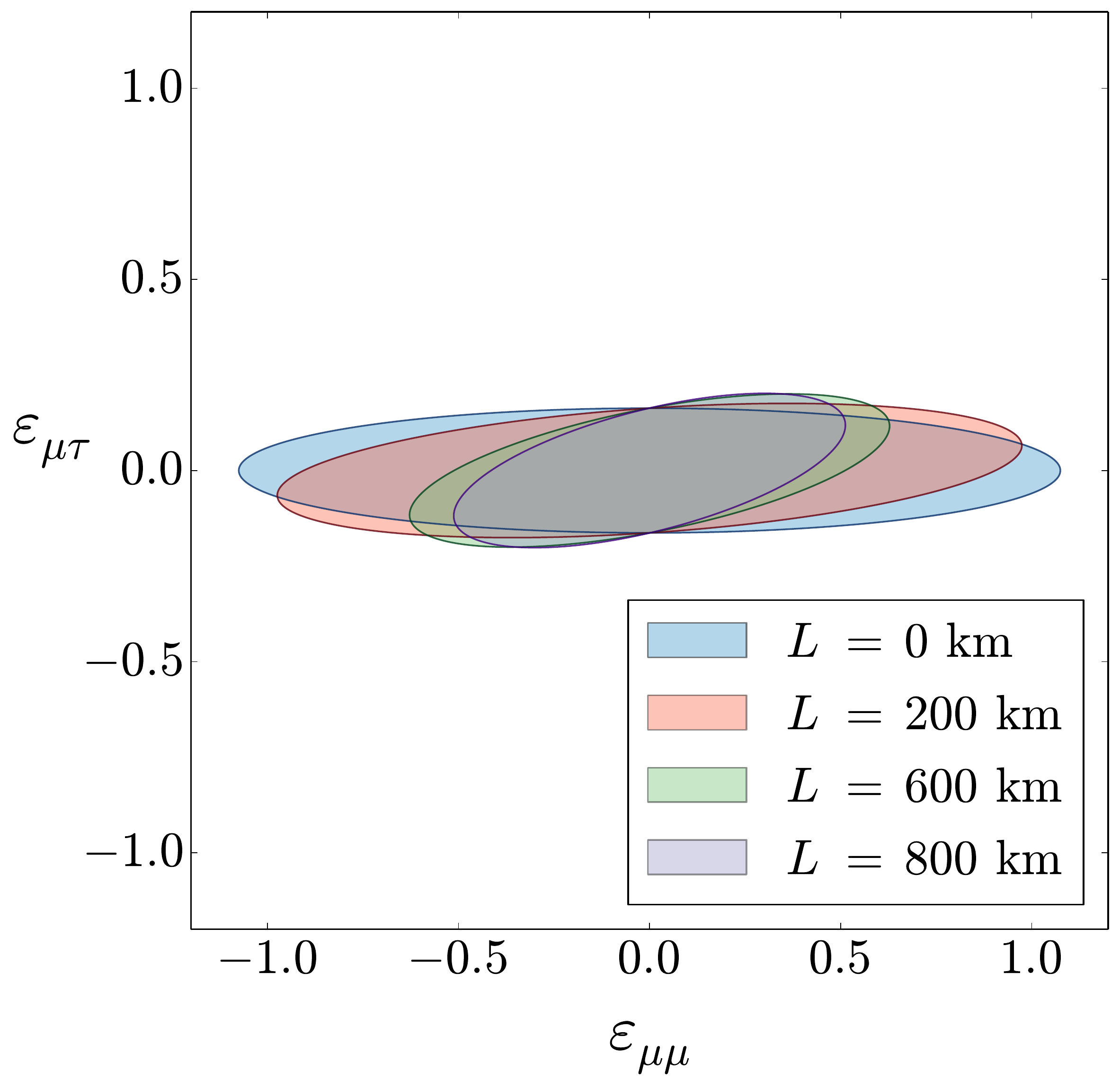}}
	\caption{Allowed regions in the $ (\varepsilon_{\mu\mu},\varepsilon_{\mu\tau}) $ parameter space for $ \eta = 0 $ (left) and $ \eta = \pi/2 $ (right) for four different values of the baseline $L$.}
\label{fig:differentLplot}
\end{figure}

While we have so far restricted our analysis to the MINOS experiment, it is interesting to consider what constraints could be made at different baselines for an experiment similar in design to MINOS. To the left of Fig. \ref{fig:differentLplot} we set $ \eta=0 $ and examine the allowed regions in the $ \varepsilon_{\mu\mu}$ -- $\varepsilon_{\mu\tau} $ space for different values of the baseline $ L $, derived for illustrative purposes from the MINOS limit $ S_{\mu\mu} \lesssim 0.026  $. We see that at zero distance this limit bounds $ |\varepsilon_{\mu\mu}|\lesssim 0.16 $, while $\varepsilon_{\mu\tau}$ remains unbounded. This is clear from the expansion of $ P_{\nu_{\mu}\rightarrow\bar\nu_{\mu}} $ -- the functions $ F^{(\mu)}_{\tau} $ and $ G^{(\mu)}_{\mu\tau} $ are directly proportional to $ \sin \Big(\eta -\frac{\delta m^{2}L}{4E_{\mathbf{q}}}\Big) $ which vanishes at $ L= 0 $. The first term in $F_{\mu}^{(\mu)}$ always contributes to $ |\varepsilon_{\mu\mu}|^{2} $ while only oscillation terms contribute to $ |\varepsilon_{\mu\tau}|^{2} $ and $ \varepsilon_{\mu\mu}\varepsilon_{\mu\tau} $. At the larger baselines of 200, 600 and 800 km the functions $ F^{(\mu)}_{\tau} $ and $ G^{(\mu)}_{\mu\tau} $ are non-zero and thus the bounded areas again become ellipses. As the $L$ increases the bounded area becomes more and more circular, improving the bound in the $ \varepsilon_{\mu\tau} $ direction. For $ \eta = \pi/2 $ shown in Fig. \ref{fig:differentLplot} (right), $ F^{(\mu)}_{\tau} $ and $ G^{(\mu)}_{\mu\tau} $ are non-zero at $ L=0 $ and so the bound is an ellipse at zero distance. Unlike for $ \eta=0 $ the bound in the $ \varepsilon_{\mu\mu} $ direction improves as $ L $ increases.

\begin{figure}[t]
	\includegraphics[width=7cm]{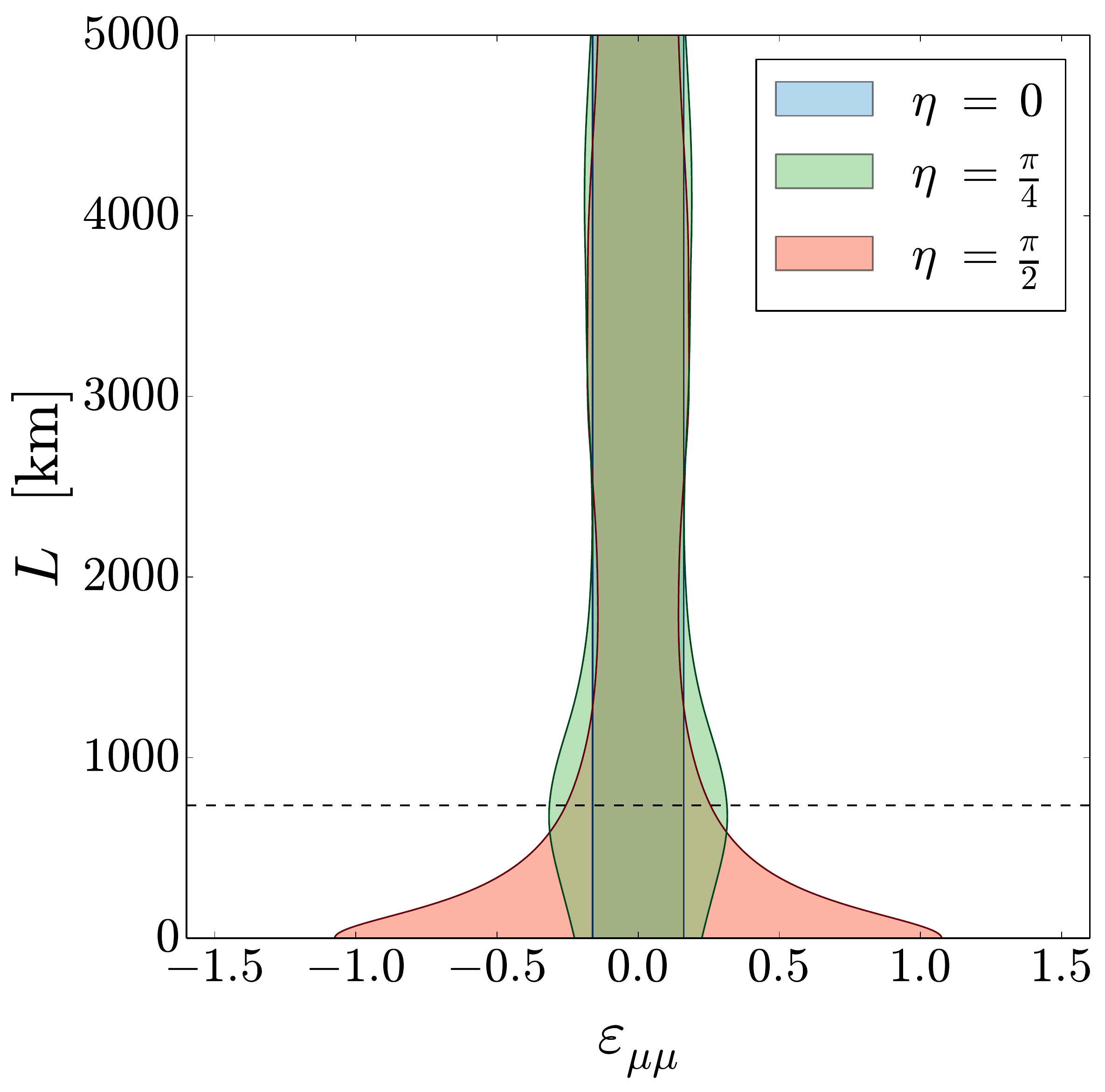}\hspace{0.5em}\includegraphics[width=7cm]{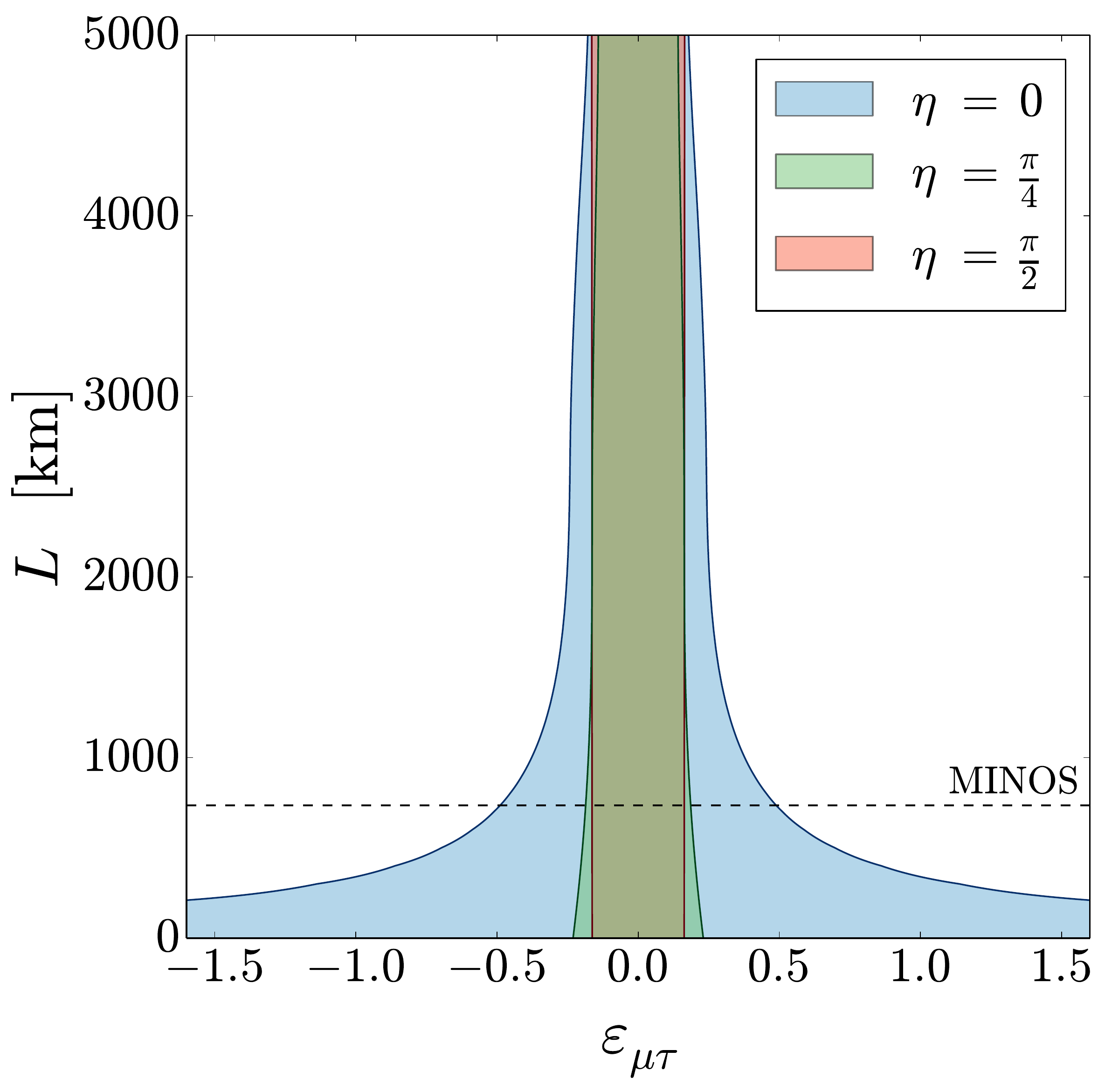}
	\caption{Constraints on $ \varepsilon_{\mu\mu} $ for $ \varepsilon_{\mu\tau} = 0 $ (left) and $ \varepsilon_{\mu\tau}$ for $ \varepsilon_{\mu\mu} = 0 $ (right) and as a function of the baseline $ L $ for three values of the Majorana phase $ \eta $. The baseline of MINOS is indicated by the dashed line.}
\label{fig:baselinefunctionplot}
\end{figure}

\begin{table}[b] 
	\begin{tabular}{l|l|l}
		\hline NSI coefficient & ~~Fixed energy upper bound ~ & ~~Integrated upper bound ~  \\ \hline\hline
		~~~~~~~$ |\varepsilon_{\mu\mu}| $ & ~~~~~~~~~~~~$ 0.11-0.76 $  & ~~~~~~~~~~~$0.15-0.55$ \\
		~~~~~~~$ |\varepsilon_{\mu\tau}| $& ~~~~~~~~~~~~~$ 0.12-\infty$ & ~~~~~~~~~~~$0.16-0.66 $ \\ \hline
	\end{tabular}
	\caption{Upper bounds from the MINOS experiment on the LNV NSI coefficients in the 2$\nu$ mixing approximation. The range indicates the best and worst upper bound depending on the choice of the Majorana phase. Middle: bounds derived at a fixed neutrino energy of 3 GeV. Right: bounds derived by integrating over the energy-dependent NuMI flux, probability and cross section.}
\label{table:ii}
\end{table}

Finally, it is interesting to look at the bounds on $ \varepsilon_{\mu\mu} $ as a function of $ L $ and for $ \varepsilon_{\mu\tau}=0 $. In Fig. \ref{fig:baselinefunctionplot} (left) we plot allowed values of $ \varepsilon_{\mu\mu}$ along the $x$-axis as a function of $ L $ along the $y$-axis. For $ \eta = 0  $ (and $\eta = n\pi$, $ n\in \mathbb{Z} $) the oscillating part of $ P_{\nu_{\alpha}\rightarrow\bar{\nu}_{\beta}} $ exactly cancels the oscillating part of $ P_{\nu_{\alpha}\rightarrow\nu_{\beta}} $, and thus the bound on $ \varepsilon_{\mu\mu} $ does not depend on $ L $. For $ \eta = \pi/2$ the constraint at zero distance is far less stringent, but decreases appreciably from 0 km up to 1000 km. For $ \eta=\pi/4 $ the constraint worsens as $ L $ reaches $\sim 800$ km but improves at larger baselines. For $ L \gtrsim$ $ 2000 $ km the constraints for non-zero $ \eta $ slowly oscillate but are roughly the same as for $ \eta=0 $, i.e. $ |\varepsilon_{\mu\mu}| \lesssim 0.15 $. We show in Fig. \ref{fig:baselinefunctionplot} (right) a similar plot for $\varepsilon_{\mu\tau}$, setting $\varepsilon_{\mu\mu}=0$ and shading the allowed regions as a function of the baseline. At zero baseline $\varepsilon_{\mu\tau}$ is unbounded for $\eta = 0$, as discussed previously. For large baseline the upper limits converge to $ |\varepsilon_{\mu\mu}| \lesssim 0.16 $. 

We summarise the constraints made in the $2\nu$ mixing approximation on the two coefficients $\varepsilon_{\mu\mu}$ and $\varepsilon_{\mu\tau}$ in Table \ref{table:ii}. Here we allow one coefficient at a time to be non-zero, computing an upper bound in the fixed energy approximation (middle) and integrating over the energy (right). The lower and upper values are the best and worst upper bounds, respectively, as the phase $\eta$ is varied. One can see that $\varepsilon_{\mu\tau}$ is unbounded for a specific value of $\eta$ when the energy is fixed. In this analysis we have of course taken the best fit values of the standard mixing parameters to be fixed. A rigorous fit to the data would need to let these parameters vary along with the $\varepsilon$ coefficients, as taken into account for LNC NSI in Refs. \cite{Blennow:2007pu,Ohlsson:2013nna,Farzan:2017xzy,Esteban:2018ppq}.

\vspace{-0.5em}
\subsection{CONSTRAINTS FROM THE MINOS EXPERIMENT IN THE THREE NEUTRINO MIXING SCHEME}
\vspace{-0.5em}

We now generalise the analysis to the 3$\nu$ mixing scheme. We use the standard parametrisation of the PMNS mixing matrix
\begin{equation}
U =
\underbrace{\begin{pmatrix}
1 & 0 & 0\\
0 & c_{23} & s_{23} \\
0 & -s_{23} & c_{23}
\end{pmatrix}}_{R_{23}}
\cdot
\underbrace{\begin{pmatrix}
c_{13} & 0 & s_{13}e^{-i\delta}\\
0 & 1 & 0 \\
-s_{13}e^{i\delta} & 0 & c_{13}
\end{pmatrix}}_{W_{13}}
\cdot
\underbrace{\begin{pmatrix}
c_{12} & s_{12} & 0\\
-s_{12} & c_{12} & 0 \\
0 & 0 & 1
\end{pmatrix}}_{R_{12}}
\cdot
\underbrace{\begin{pmatrix}
1 & 0 & 0 \\
0 & e^{i\frac{\alpha_{2}}{2}} & 0 \\
0 & 0 & e^{i\frac{\alpha_{3}}{2}} 
\end{pmatrix}}_{D}~,
\end{equation}
where $s_{ij}=\sin\theta_{ij}$, $c_{ij}=\cos\theta_{ij}$ and ($\alpha_2$, $\alpha_3$) are Majorana phases. We now look to probe the $3 \times 3$ flavour structure of the NSI coefficient matrix $\varepsilon_{\beta\alpha}$ in which all elements are taken to be real. We again expand the effective non-standard oscillation probability as in Eqs. (\ref{eq:expand}) and (\ref{eq:fandg}),
\begin{equation} \label{eq:expand2}
\begin{split}
\begin{aligned}
P_{\nu_{\alpha}\rightarrow\bar{\nu}_{\beta}}&\approx\bigg|\sum\limits^{3}_{i}U_{\alpha i}^{*}~\gamma_{\beta i}~e^{-i\frac{m_{i}^{2} }{2E_{\mathbf{q}}}L}\bigg|^{2}\\
&=\sum_{\lambda=e,\mu,\tau}F^{(\alpha)}_{\lambda}(L, E_{\mathbf{q}},\boldsymbol{\zeta},\alpha_{2},\alpha_{3})~\varepsilon_{\beta\lambda}^{2} +\sum\limits_{\lambda<\lambda'=e,\mu,\tau} G^{(\alpha)}_{\lambda\lambda'}(L, E_{\mathbf{q}},\boldsymbol{\zeta},\alpha_{2},\alpha_{3}) ~\varepsilon_{\beta\lambda}\varepsilon_{\beta\lambda'}~,
\end{aligned}
\end{split}
\end{equation}
where $F_{\lambda}^{(\alpha)}$ and $G_{\lambda\lambda'}^{(\alpha)}$ are now complicated functions of the baseline $L$, neutrino energy $E_{\mathbf{q}}$, neutrino mixing parameters $\boldsymbol{\zeta}$ and Majorana phases $\alpha_2$ and $\alpha_3$. Using the best fit values for the mixing parameters $\theta_{12}$,  $\theta_{23}$,  $\theta_{13}$,  $\Delta m^2_{12}$ and $\Delta m^2_{23}$ and setting $\delta $ to the value hinted at by the most recent data (again in the NO scheme), we numerically rotate to the flavour space coefficients as performed in the  second line of Eq. (\ref{eq:expand2}) to find $F_{\lambda}^{(\alpha)}$ and $G_{\lambda\lambda'}^{(\alpha)}$ as functions of $L$, $E_{\mathbf{q}}$, $\alpha_2$ and $\alpha_3$. It is useful to compare the two Majorana phases used here to the single phase in the $2\nu$ mixing approximation. Comparing the numerical expression of $F^{(\mu)}_{\mu}$ in the $3\nu$ scheme to that in Eq. (\ref{eq:functions}) for the $2\nu$ scheme and taking the limits $\Delta m^2_{12}\rightarrow 0$, $\Delta m^2_{13}\rightarrow \Delta m^2_{23} $, we find that $\vartheta \approx \theta_{23}$, $\delta m^2 \approx \Delta m^2_{23}$ which we had already assumed in Sec. IV A, and that $\eta\approx(\alpha_3-\alpha_2)/2$.

\begin{figure}[t]
	\hspace{-1em}\includegraphics[width=5.6cm]{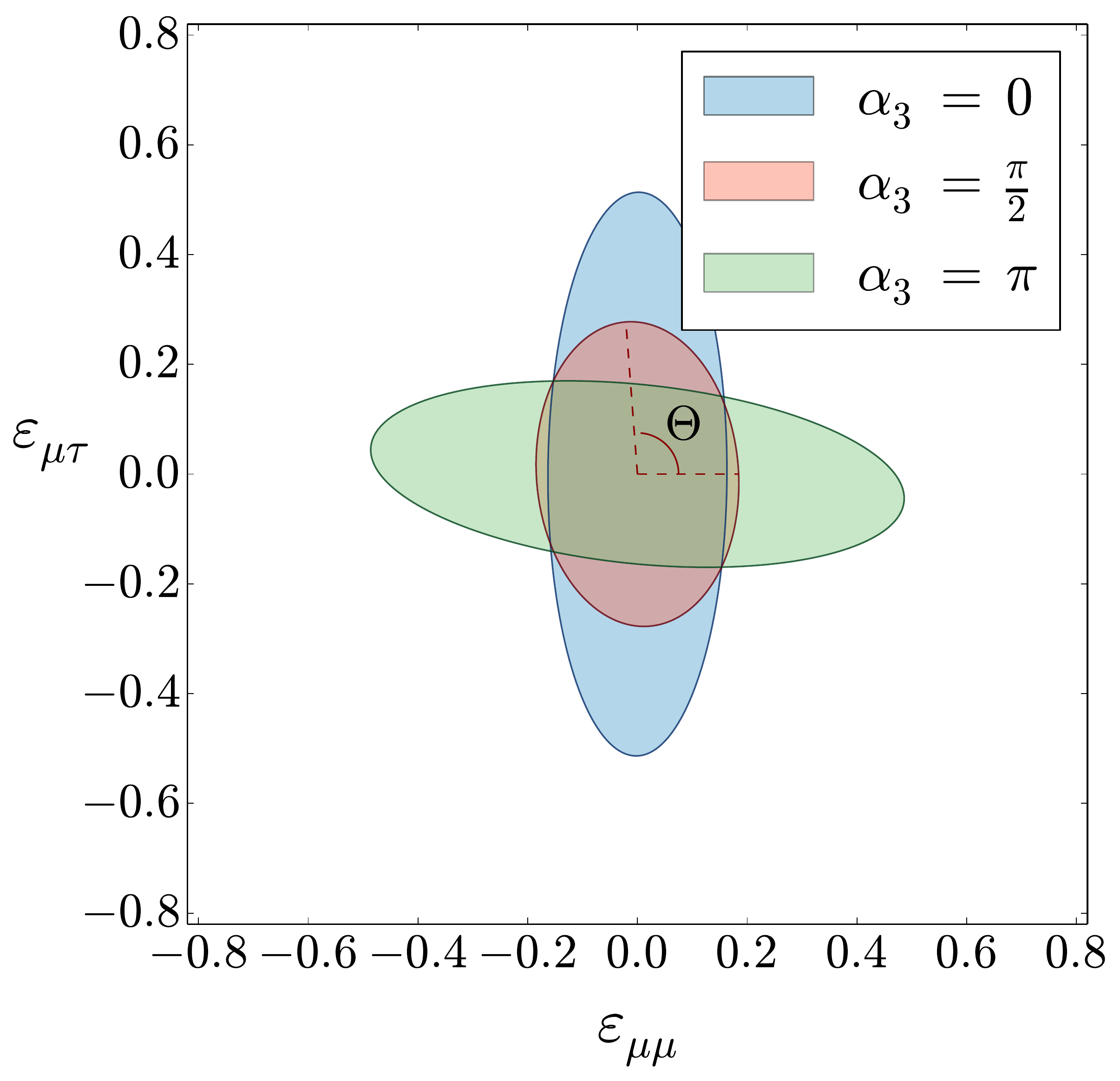}\includegraphics[width=5.6cm]{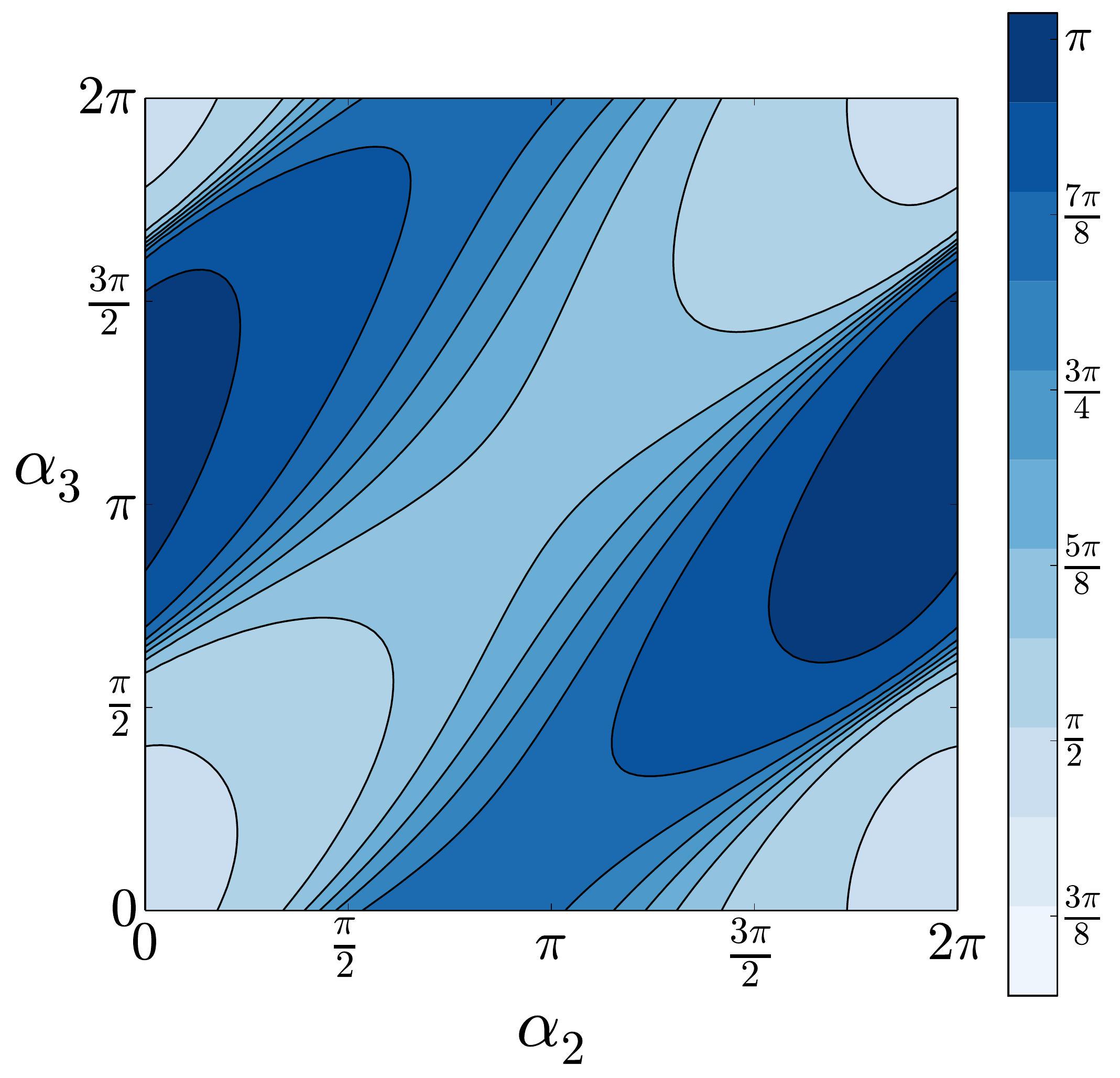}\hspace{-0.2em}\includegraphics[width=5.6cm]{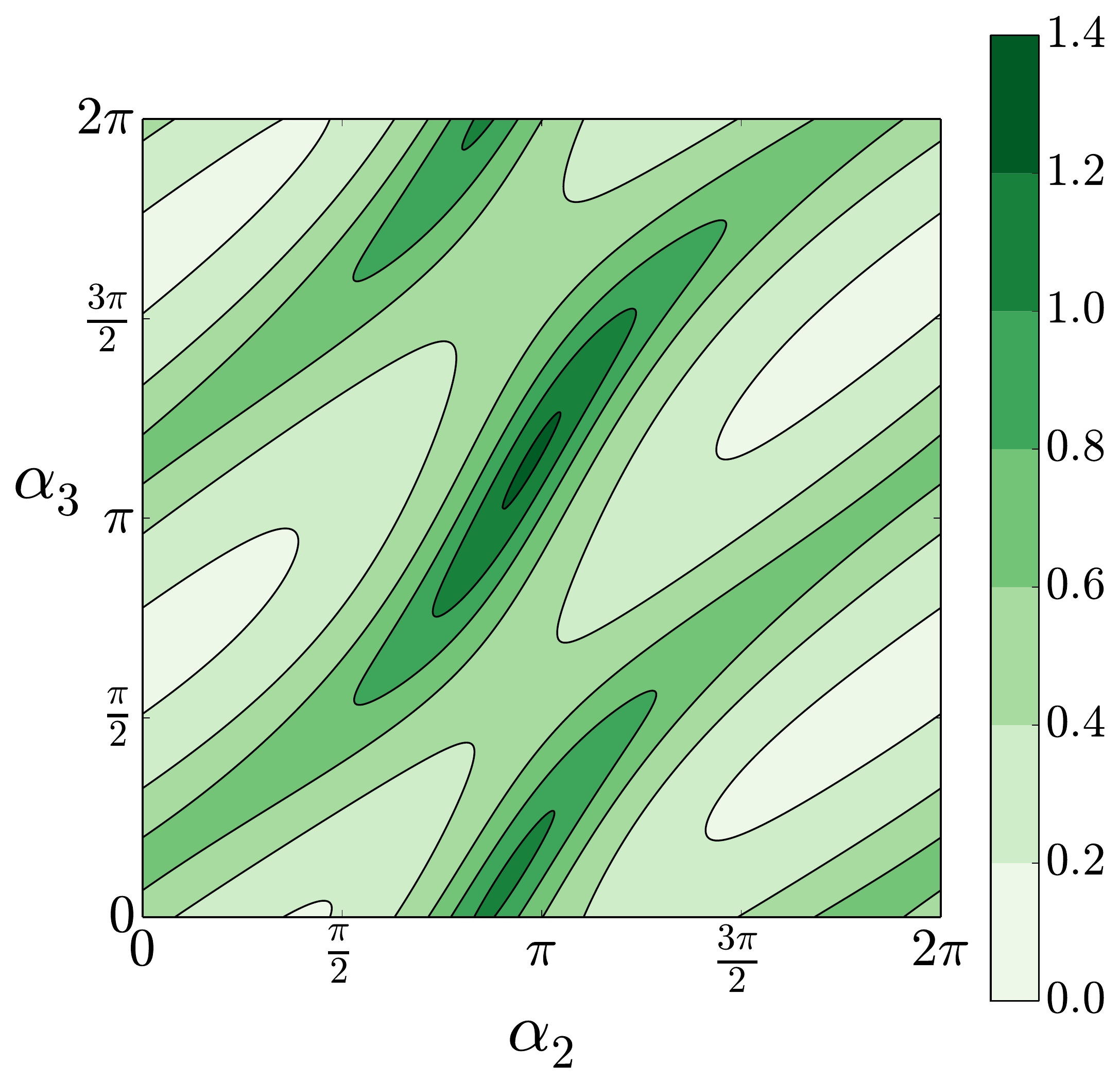}
	\caption{Left: allowed regions in the $(\varepsilon_{\mu \mu},\varepsilon_{\mu \tau})$ parameter space for $\varepsilon_{\mu e} = 0$, $L$ = 735 km, $\alpha_2 = 0 $ and three values of the Majorana phase $\alpha_3$. Middle: angle $ \Theta $ from the positive $\varepsilon_{\mu\mu}$ axis to the semi-major axis of the constraint ellipse as a function of the phases. Right: eccentricities of the constraint ellipse also as a function of the phases.}
\label{fig:MINOS3nu}
\end{figure}

We now return to the interpretation of the MINOS bound $S_{\mu\mu}\lesssim 0.026 $. This ratio is defined in the same manner as Eq. (\ref{eq:signal}) -- multiplying Eq. (\ref{eq:expand2}) by the differential production flux and detection cross section and integrating over the NuMI beam energy between 0 and 20 GeV. Now in the 3$\nu$ mixing scheme, $S_{\mu\mu} < 0.026 $ can be converted into an allowed region in the $(\varepsilon_{\mu e},\varepsilon_{\mu \mu},\varepsilon_{\mu \tau})$ parameter space which depends on the value of the Majorana phases $\alpha_{2} $ and $ \alpha_{3}$. Firstly, and in order to compare with bounds in the 2$\nu$ mixing approximation, we set $\varepsilon_{\mu e} = 0 $ and depict in Fig. \ref{fig:MINOS3nu} (left) the allowed regions in the $(\varepsilon_{\mu \mu},\varepsilon_{\mu \tau})$ parameter space for $\alpha_2=0$ and three different values of $\alpha_3$. The ellipses are of similar size to those for the 2$\nu$ mixing scheme but have shapes and orientations (which we define as the ellipse eccentricity $e$ and anticlockwise angle $\Theta$ from the positive $\varepsilon_{\mu\mu}$ axis to the semi-major axis, respectively) with similar dependences on the Majorana phases. We show these dependences as contour plots in Fig. \ref{fig:MINOS3nu} (middle and right). We can see that the dependence of the angle $\Theta$ is approximately the same as $\eta \approx (\alpha_3-\alpha_2)/2$, which is evident from the diagonal lines of roughly constant $\Theta$ along $\alpha_3=\alpha_2+C$. For $\alpha_2 = \alpha_3 = 0$ for example, we see that the constraint is better in the $\varepsilon_{\mu\mu}$ direction -- this is similar to $\eta = 0$ in the $2\nu$ scheme. Likewise, for $\alpha_2 = 0,~ \alpha_3 = \pi$, the constraint is better in the $\varepsilon_{\mu\tau}$ direction which is similar to $\eta = \pi$. We see that the largest eccentricity  occurs at $\alpha_2\approx\alpha_3\approx \pi$ -- this coincides with the semi-major axis pointing in the $\varepsilon_{\mu\tau}$ direction and therefore the upper bound on $|\varepsilon_{\mu\tau}|$ can be slightly larger than the upper bound on $|\varepsilon_{\mu\mu}|$.

\begin{figure}[t]
	{\hspace{-0.5em}\includegraphics[width=5.45cm]{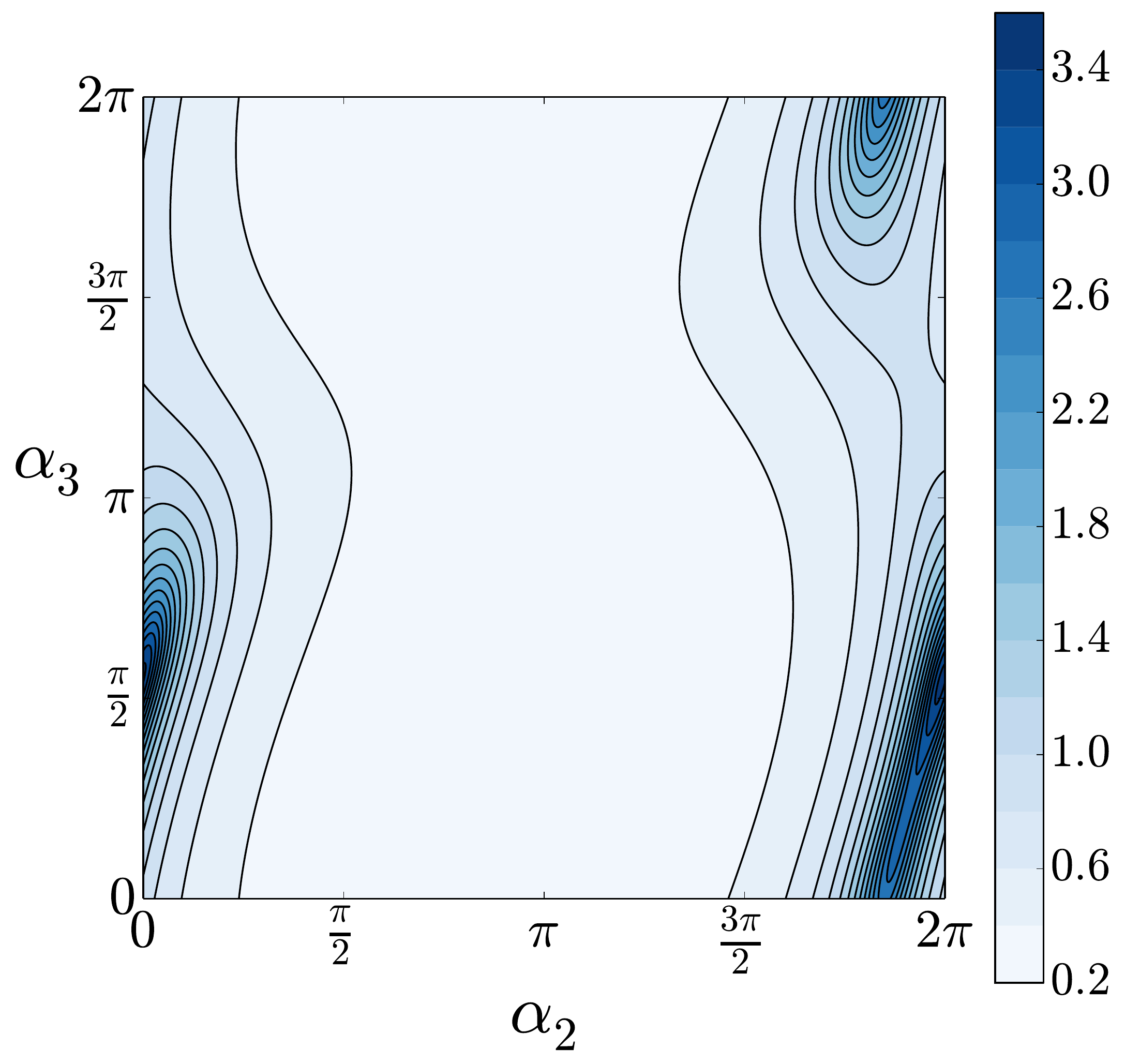}\hspace{0em}\includegraphics[width=5.55cm]{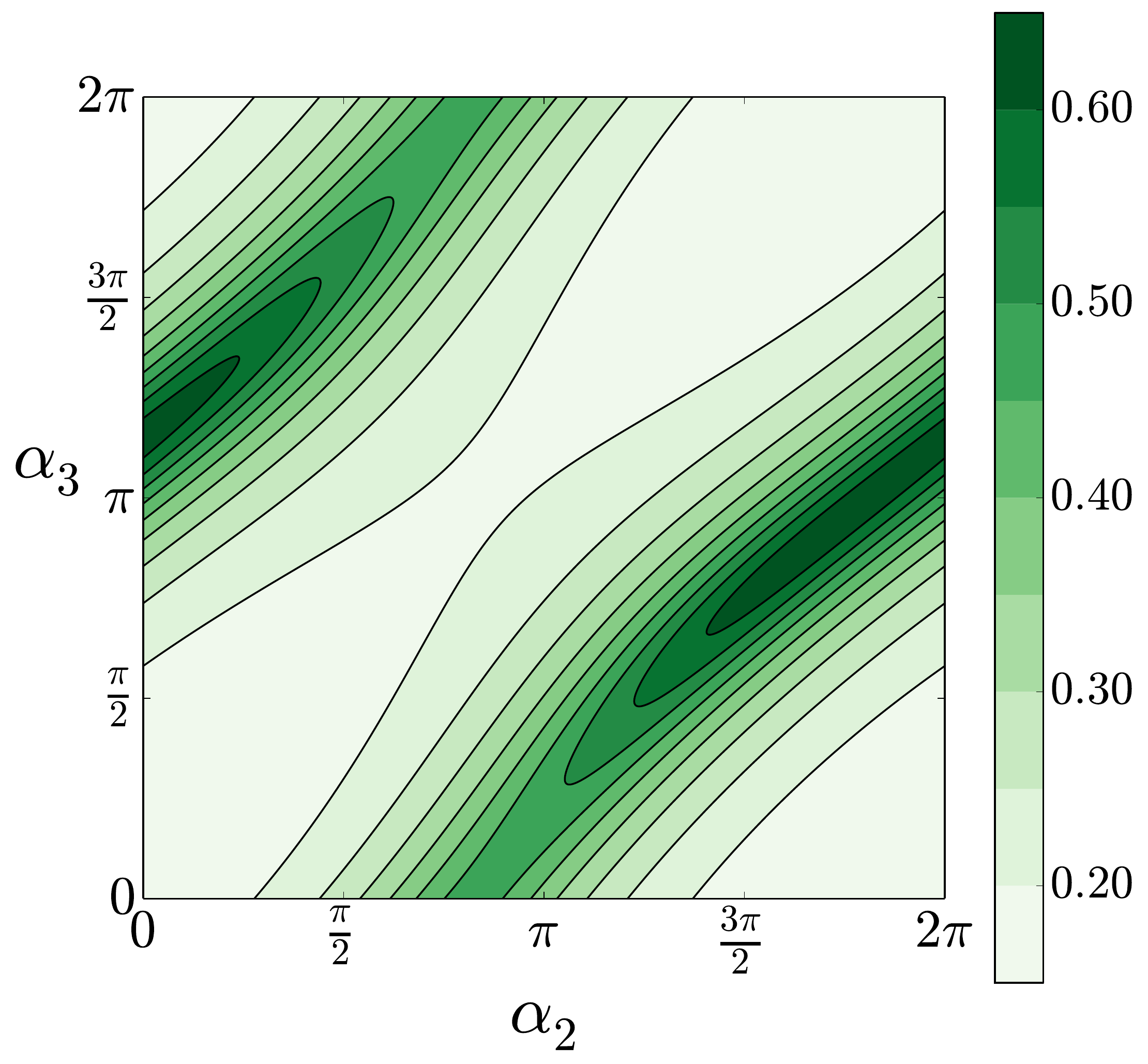}\hspace{-0em}}\includegraphics[width=5.55cm]{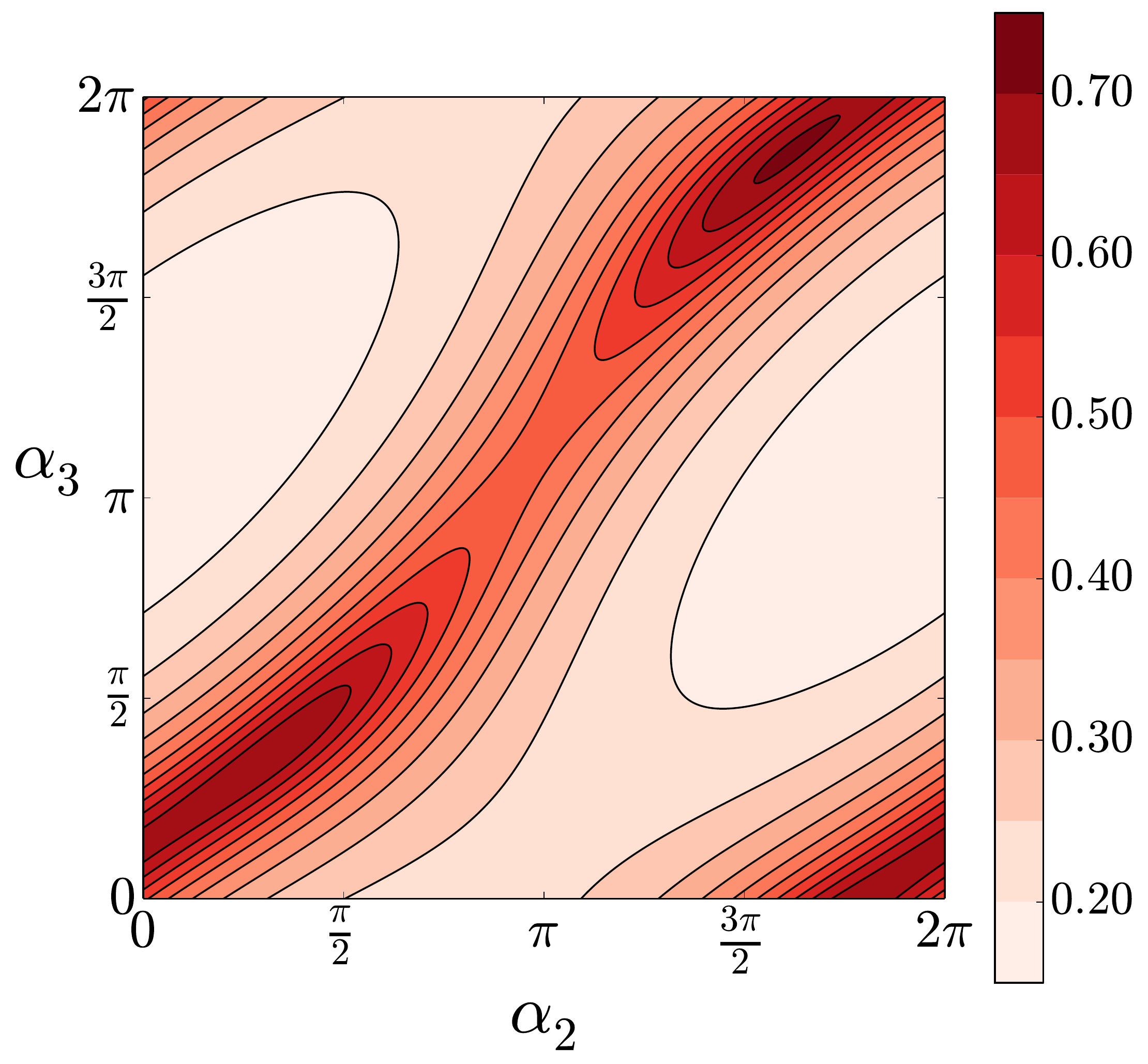}
	\caption{Upper bounds on the NSI coefficients $\varepsilon_{\mu e}$ (left), $\varepsilon_{\mu\mu}$ (centre) and $\varepsilon_{\mu\tau}$ (right), as a function of the Majorana phases $\alpha_2$ and $\alpha_3$, derived from the MINOS limit $S_{\mu\mu} < 0.026 $. Best fit values for the mixing parameters $\theta_{12}$, $\theta_{13}$ and $\theta_{23}$, $\Delta m_{12}^2$ and $\Delta m_{23}^2$ and $\delta$ are taken in the NO.}
	\label{fig:musectorMINOS}
\end{figure}

In Fig. \ref{fig:musectorMINOS} we plot the upper bounds on each NSI coefficient in the $\mu$ sector as a function of the Majorana phases when taking each to be non-zero at a time. We can see the values of $\alpha_2$ and $\alpha_3$ where the upper bounds are more or less stringent. For $\varepsilon_{\mu e}$ (left) we can see that for most of the parameter space the upper bound is of order  $|\varepsilon_{\mu e}|\lesssim 1$, but worsens for particular values of the phases to $|\varepsilon_{\mu e}|\lesssim 3.4$. Upper bounds for the other two coefficients are in the ranges $|\varepsilon_{\mu\mu}|\lesssim 0.2 - 0.6$ and $|\varepsilon_{\mu\tau}|\lesssim 0.2 - 0.7$. We summarise these constraints, along with the results from the following sections, to the right of Table \ref{table:iii}. The lower and upper values are the \textquoteleft best' and \textquoteleft worst' upper bounds depending on the value of the Majorana phases $\alpha_2$ and $\alpha_3$.

While we have so far considered only MINOS, OPERA was another LBL oscillation experiment to employ a magnetic field in the far detector \cite{Acquafredda:2009zz,Galati:2016zqe}. Unlike MINOS, OPERA measured neutrinos from the CNGS beam at CERN that were above the threshold for $ \tau^{\pm} $ production. The main physics goal of the experiment was to confirm $ \nu_{\tau} $ appearance -- around 10 $ \tau^{\pm} $ events were recorded over four years of data taking \cite{Agafonova:2018auq}. Unfortunately the experiment was only able to distinguish the charge of one $ \tau^{-} $ event at 5$ \sigma $ significance, while the other charges were undetermined. The statistics are therefore too low to comment on OPERA's sensitivity to lepton number. A future high-statistics LBL experiment above the $\tau^{\pm}$ threshold like OPERA would therefore be able to probe the $\tau$ sector NSI coefficients, $ \varepsilon_{\tau e} $, $ \varepsilon_{\tau\mu} $ and $ \varepsilon_{\tau\tau}$.

\vspace{-0.5em}
\subsection{CONSTRAINTS FROM THE KAMLAND EXPERIMENT IN THE THREE NEUTRINO MIXING SCHEME}
\vspace{-0.5em}

Operating for a window of 185.5 days between March 4 and December 1, 2002, the LBL reactor neutrino experiment KamLAND conducted a search for solar $\bar{\nu}_e $ with the characteristic flux of $^8$B $\nu_e$. The analysis of Ref. \cite{Eguchi:2003gg} hence assumed the $\bar{\nu}_e$ of interest to be descendant from the solar $\nu_e$, either through spin precession in the Sun's magnetic field due to a non-zero neutrino transition magnetic moment or neutrino decay. Hence the bound derived on the ratio $S_{ee} \lesssim 2.8\times 10^{-4}$ at 90 \% C.L. can be used in a similar fashion to MINOS to place constraints on the NSI coefficients.

If the initial $\nu_e$ are produced from the beta decay of $^8$B and propagate from the solar core to the solar surface, and then through the vacuum to the KamLAND detector. The oscillation probability must therefore take into account the resonant conversion of solar $\nu_e$ to $\nu_{\mu}$ and $\nu_{\tau}$ through the Mikheyev-Smirnov-Wolfenstein (MSW) effect, a consequence of the slowly decreasing matter potential from the Sun's core to surface. We make the approximation that the conversion is adiabatic and utilise the $\Delta m_{12}^2\ll \Delta m_{23}^2 $ hierarchy to write the standard $\nu_{\alpha}\rightarrow\nu_{\beta}$ conversion probability in a similar form to that in Ref. \cite{Esteban:2018ppq},
\begin{equation} \label{eq:msw}
\begin{split}
\begin{aligned}
P^{\mathrm{eff}}_{\nu_{\alpha} \rightarrow \nu_{\beta}} \approx\left|\sum_{i, j=1,2}^{3}\left(R^{23} W^{13}\right)_{\alpha i}^{*}\left(R^{23} W^{13}\right)_{\beta j} \mathcal{U}_{i j}(x)\right|^{2}+\left|U_{\alpha 3}\right|^{2}\left|U_{\beta 3}\right|^{2}~,
\end{aligned}
\end{split}
\end{equation}
where $R^{23}$ and $W^{13}$ are Euler rotations making up the standard parametrisation of the PMNS matrix and $\mathcal{U}$ is a $2\times 2$ unitary matrix satisfying
\begin{equation}
i \frac{d}{d x} ~\mathcal{U}(x)=\frac{\widehat{\mathrm{M}}_{2 \times 2}^{2}}{2 E_{\mathrm{q}}}~ \mathcal{U}(x)~.
\end{equation}
$\widehat{\mathrm{M}}^2_{2\times2}$ is the effective $2 \times 2$ squared mass matrix
\begin{equation}
\widehat{\mathrm{M}}_{2 \times 2}^{2}=\frac{\Delta m_{12}^{2}+c_{13}^{2} A_{\mathrm{CC}}}{2}+\frac{1}{2} \left( \begin{array}{cc}{-\cos 2 \theta_{12} \Delta m_{12}^{2}+c_{13}^{2} A_{\mathrm{CC}}} & {\sin 2 \theta_{12} \Delta m_{12}^{2}} \\ {\sin 2 \theta_{12} \Delta m_{12}^{2}} & {\cos 2 \theta_{12} \Delta m_{12}^{2}-c_{13}^{2} A_{\mathrm{CC}}}\end{array}\right)~,
\end{equation}
where $A_{\mathrm{CC}} = 2\sqrt{2}G_{F}E_{\mathbf{q}}N_{e}$ and $N_e$ is the electron density in the Sun. In order to construct $S_{ee}$ we require the non-standard oscillation equivalent of Eq. (\ref{eq:msw}). This can be derived from Eq. (\ref{eq:expand}), but an exact formula taking into account the MSW effect, even in the $\Delta m_{12}^2\ll \Delta m_{23}^2 $ limit, is beyond the scope of this work. The possibility that the NSI occurs at production would also complicate the analysis, because $\bar{\nu}_e$ would experience a different matter effect while propagating through the Sun -- we therefore concentrate on the NSI being at detection. 

\begin{table}[t]
	\begin{tabular}{l|l|l|l|l}
		\hline~NSI coefficient~ &  ~~Previous upper bound~ & ~~~~~Process &~LBL upper bound~  & ~LBL experiment \\ \hline\hline
		~~~~~~~$ |\varepsilon_{ee}| $ & ~~$2.1\times 10^{-9} -6.3\times 10^{-9} $~ & & ~~~~~~~~~~$ 0.017 $~ &    \\
		~~~~~~~$ |\varepsilon_{e\mu}| $& ~~~~~~$2.9 \times 10^{-9}-\infty $~ &~~$ 0\nu\beta\beta  $ ($ ^{76}\mathrm{Ge} $)~~ & ~~~~~~~~~~$ 0.017 $  &  ~~~~~KamLAND \\
		~~~~~~~$ |\varepsilon_{e\tau}| $& ~~~~~~$2.6 \times 10^{-9}-\infty $ & & ~~~~~~~~~~$ 0.015 $ &  \\
		\hline
		~~~~~~~$ |\varepsilon_{\mu e}| $& ~~~$ \sim 4\times 10^{3}-1\times 10^{4} $ & & ~~~~~~$0.22-3.47$ &   \\
		~~~~~~~$ |\varepsilon_{\mu\mu}| $& ~~~~~~$\sim6\times 10^{3}-\infty$ ~& ~~~~~$\mu^- - e^+$~ & ~~~~~~$0.16 -0.63 $  & ~~~~~~~MINOS  \\
		~~~~~~~$ |\varepsilon_{\mu\tau}| $& ~~~~~~$\sim 5\times 10^{3}-\infty$ ~ & & ~~~~~~$0.16 -0.71 $ &  
		\\ \hline
	\end{tabular}
	\caption{Upper bounds on the LNV NSI flavour coefficients in the $e$ and $\mu$ sectors. Left: bounds derived from conventional microscopic LNV processes, with $0\nu\beta\beta$ decay being the most effective the $e$ sector and $\mu^{-}- e^{+}$ conversion loosely constraining the $\mu$ sector. Right: bounds from LBL oscillation experiments MINOS and KamLAND. Two values indicate the variation in the upper bound as $(\alpha_2,\alpha_3)$ are varied.}
	\label{table:iii}
\end{table}

We can safely assume that, by the time the solar neutrinos reach Earth, they make up an incoherent mixture of flavour eigenstates. This effectively washes out any dependence on the Majorana phases, and we can approximate the non-standard oscillation probability as being the effective standard probability $P^{\mathrm{eff}}_{\nu_{e}\rightarrow\nu_{\beta}}$ (taking into account the MSW effect) multiplied by the NSI coefficient $|\varepsilon_{e\beta}|^2$. This takes into account for example the oscillation $\nu_e \rightarrow\nu_{\mu,\tau}$ and then the LNV NSI process $\nu_{\mu,\tau}p\rightarrow e^+ n$ at detection. The signal ratio $S_{ee}$ and the limit set on it by KamLAND now becomes
\begin{equation} \label{eq:signal2}
S_{ee}\approx\frac{\int dE_{\mathbf{q}}~\sum\limits_{\beta}\frac{d\Gamma^{}_{\nu_{e}}}{dE_{\mathbf{q}}}\cdot P^{\mathrm{eff}}_{\nu_{e}\rightarrow\nu_{\beta}}\cdot\varepsilon_{e\beta}^2\cdot \sigma^{}_{\bar{\nu}_{\beta}} }{\int dE_{\mathbf{q}}~\frac{d\Gamma^{}_{\nu_{e}}}{dE_{\mathbf{q}}}\cdot P^{\mathrm{eff}}_{\nu_{e}\rightarrow\nu_{e}}\cdot\sigma^{}_{\nu_{e}}
}\lesssim 2.8\times 10^{-4}~.
\end{equation}
We now use Eq.~\eqref{eq:signal2} to set limits on NSI coefficients in the $e$ sector, $\varepsilon_{ee}$, $\varepsilon_{e\mu}$ and $\varepsilon_{e\tau}$. Using the $^{8}$B flux predicted by the solar model of Ref. {\cite{PhysRevC.54.411}}, we integrate the differential flux, oscillation probability and low energy cross section (which we assume to scale as $E_{\mathbf{q}}^2$) over 0.02 MeV bins in the range 8.3 $-$ 14.8 MeV, for both the numerator and the denominator. 

Taking each LNV NSI coefficient to be non-zero at a time, we show the derived upper bounds in Table \ref{table:iii}. Because there is no dependence on the Majorana phases, Eq. (\ref{eq:signal2}) sets a single possible upper bound on each coefficient. For $\varepsilon_{ee}$ for example, the numerator and denominator are identical except for the factor of $|\varepsilon_{ee}|^2$ in the numerator -- the upper bound on $|\varepsilon_{ee}|$ is therefore the square root of $2.8\times 10^{-4}$. The upper bounds on $|\varepsilon_{e\mu}|$ and $|\varepsilon_{e\tau}|$ are slightly different because the numerators of $S_{ee}$ contain different oscillation probabilities from the denominator.

\vspace{-0.5em}
\subsection{COMPARISON WITH CONSTRAINTS FROM OTHER LEPTON NUMBER VIOLATING PROCESSES}
\vspace{-0.5em}

We will now briefly compare the constraints from the LBL experiments MINOS and KamLAND to those from conventional searches for LNV processes. Neutrinoless double beta ($0\nu\beta\beta$) decay continues to be the most promising method to verify the Majorana nature of the light neutrinos. It is a highly useful probe because the black-box theorem ensures that any positive signal of $0\nu\beta\beta$ decay confirms the Majorana case even when the standard light neutrino exchange process is not the dominant mechanism \cite{Schechter:1981bd,Hirsch:2006yk,Duerr:2011zd}. An extensive part of the literature has studied non-standard mechanisms, including the LNV dimension-six operator considered in this work
\cite{Hirsch:1996qw,Deppisch:2012nb,Huang:2013kma,Geng:2016auy}. Here we simply extend this to the 3 $\times$ 3 flavour structure of the $\varepsilon$ coefficient in order to compare with the MINOS and KamLAND upper bounds.

\begin{figure}[t]
	{\hspace{-1.1em}\includegraphics[width=5.5cm]{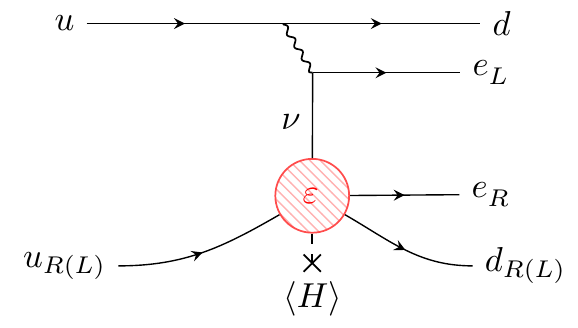}
		\includegraphics[width=5.5cm]{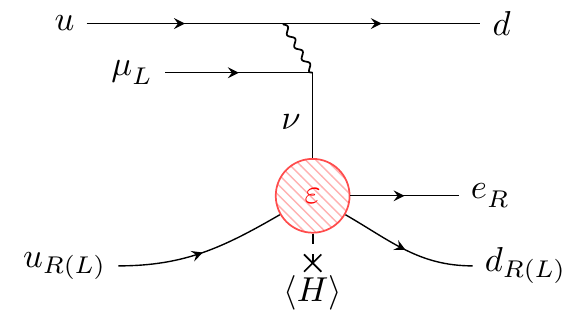}\hspace{-1em}
		\includegraphics[width=6.0cm]{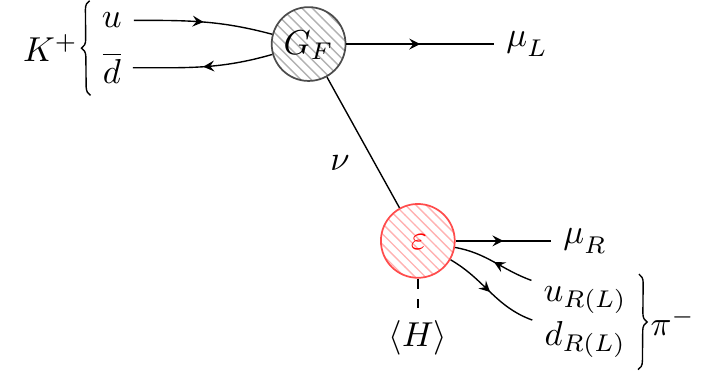}\hspace{-1em}}
	\caption{Other processes capable of probing LNV NSI coefficients. Left: $0\nu\beta\beta$ decay. Middle: $\mu^--e^+$ conversion. Right: kaon decay $K^+\rightarrow\pi^-\mu^+\mu^+$.}
\label{fig:LNVprocesses}
\end{figure}

Using from Ref. \cite{Muto:1989cd} the general expression for the $0\nu\beta\beta$ decay inverse half-life when a RH interaction is present at one of the interaction vertices (as shown in Fig. \ref{fig:LNVprocesses}):
\begin{equation}
\begin{split}
\begin{aligned}
[T_{1/2}^{0\nu\beta\beta}]^{-1} &=  C_{mm}~\frac{|\sum U_{e i}^2m_i|^2}{m^2_e} + C_{\gamma\gamma}\Big|\sum\limits_{i}^{3} U_{e i}\gamma^*_{e i}\Big|^2+C_{m\gamma}\mathrm{Re}\bigg[\sum\limits_{i,j}^{3} U^2_{ei} m_i U_{e j}^*\gamma_{e j}\bigg]~,
\end{aligned}
\end{split}
\end{equation}
where $C_{mm} = 1.12\times 10^{-13}$, $C_{\gamma\gamma}=4.44\times 10^{-9}$ and $C_{m\gamma}=2.19\times 10^{-11}$ are phase space and nuclear matrix element factors given by \cite{Muto:1989cd}. The inverse half-life, in a similar manner to the oscillation probability, can be expanded as
\begin{equation} \label{eq:onubbexpand}
\begin{split}
\begin{aligned}
[T_{1/2}^{0\nu\beta\beta}]^{-1} &= X(\boldsymbol{\zeta},\alpha_2,\alpha_3,m_0)+\sum\limits_{\lambda} Y_{\lambda}(\boldsymbol{\zeta},\alpha_2, \alpha_3,m_0)~\varepsilon_{e\lambda}  \\ 
& ~~~ +\sum_{\lambda}F^{(e)}_{\lambda}(\boldsymbol{\zeta},\alpha_2,\alpha_3)~\varepsilon_{e\lambda}^2+\sum\limits_{\lambda<\lambda'}G^{(e)}_{\lambda\lambda'}(\boldsymbol{\zeta},\alpha_2,\alpha_3)~\varepsilon_{e\lambda}\varepsilon_{e\lambda'}~,
\end{aligned}
\end{split}
\end{equation}
where $F^{(e)}_{\lambda}(\boldsymbol{\zeta},\alpha_2,\alpha_3)$ and $G^{(e)}_{\lambda\lambda'}(\boldsymbol{\zeta},\alpha_2,\alpha_3)$ are the functions in Eq. (\ref{eq:expand2}) at zero distance. The contribution from light neutrino exchange is $X(\boldsymbol{\zeta},\alpha_2,\alpha_3, m_0)$ which is a function of the Majorana phases and lightest neutrino mass. $Y_{\lambda}(\boldsymbol{\zeta},\alpha_2,\alpha_3, m_0)$ is the contribution from interference between the light neutrino exchange and the non-standard mechanism towards terms linear in the $\varepsilon$ coefficients. 

\begin{figure}[t]
	{\hspace{-0.5em}\includegraphics[width=5.3cm]{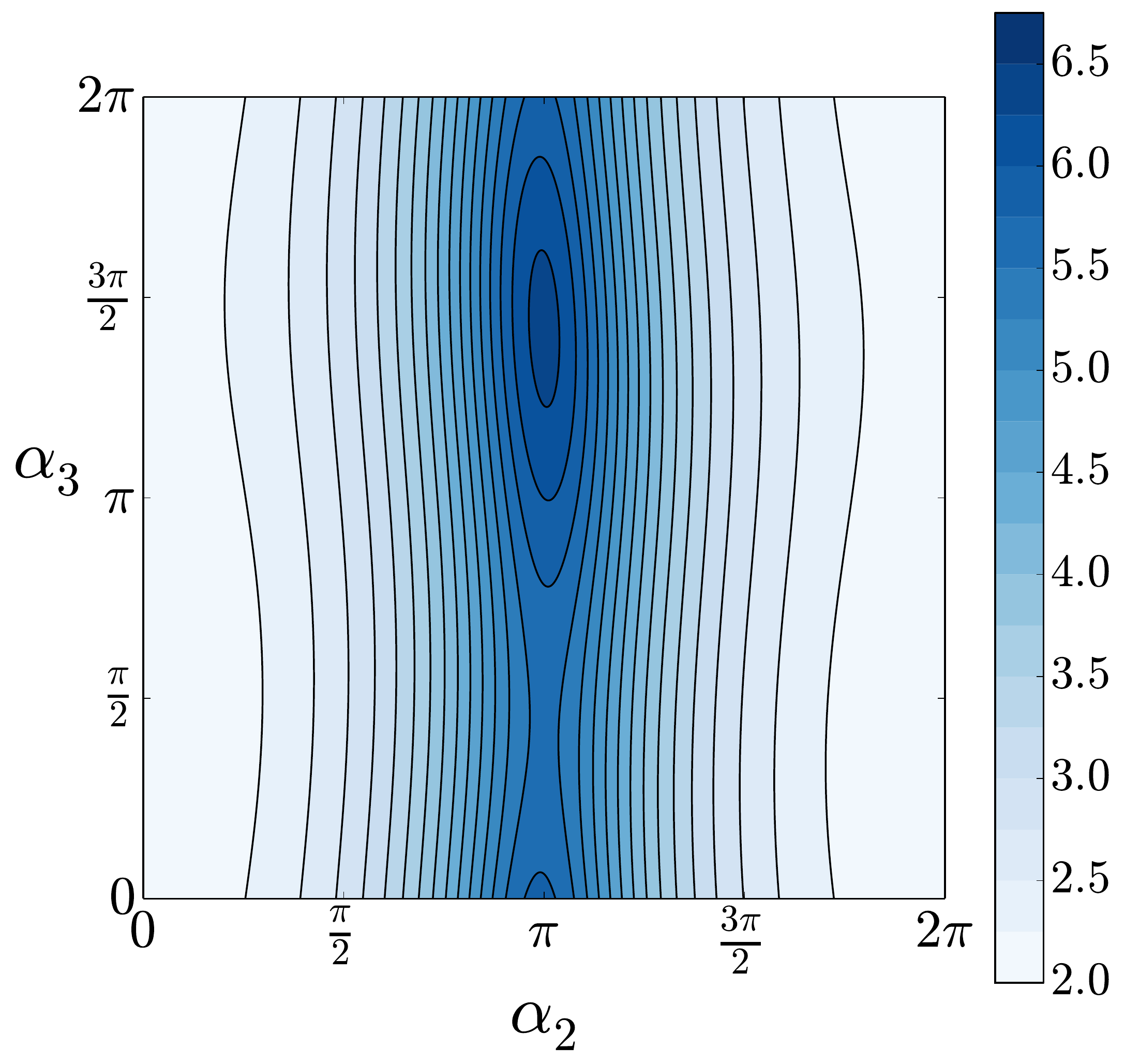}\hspace{0em}\includegraphics[width=5.7cm]{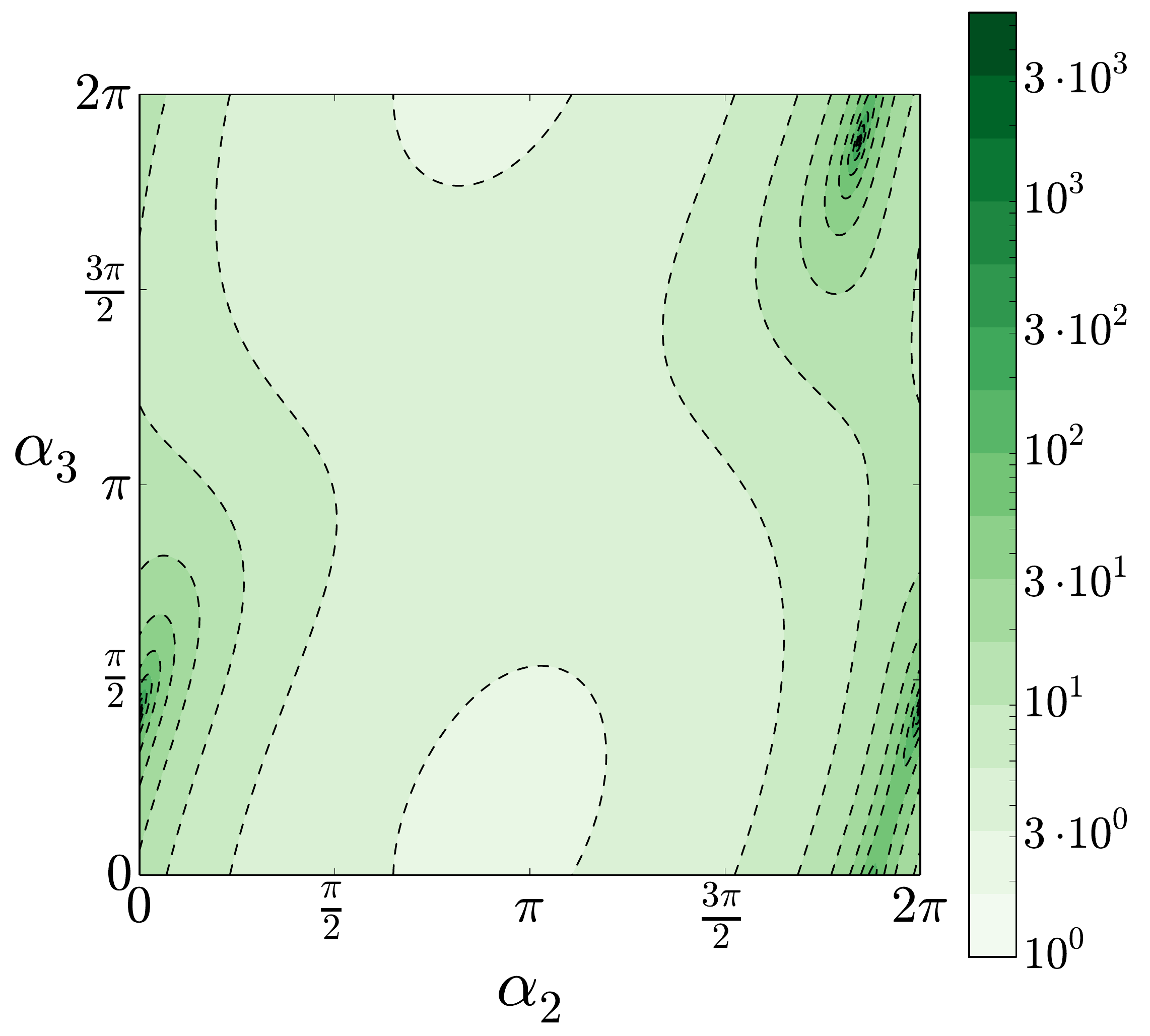}\hspace{-0em}}\includegraphics[width=5.6cm]{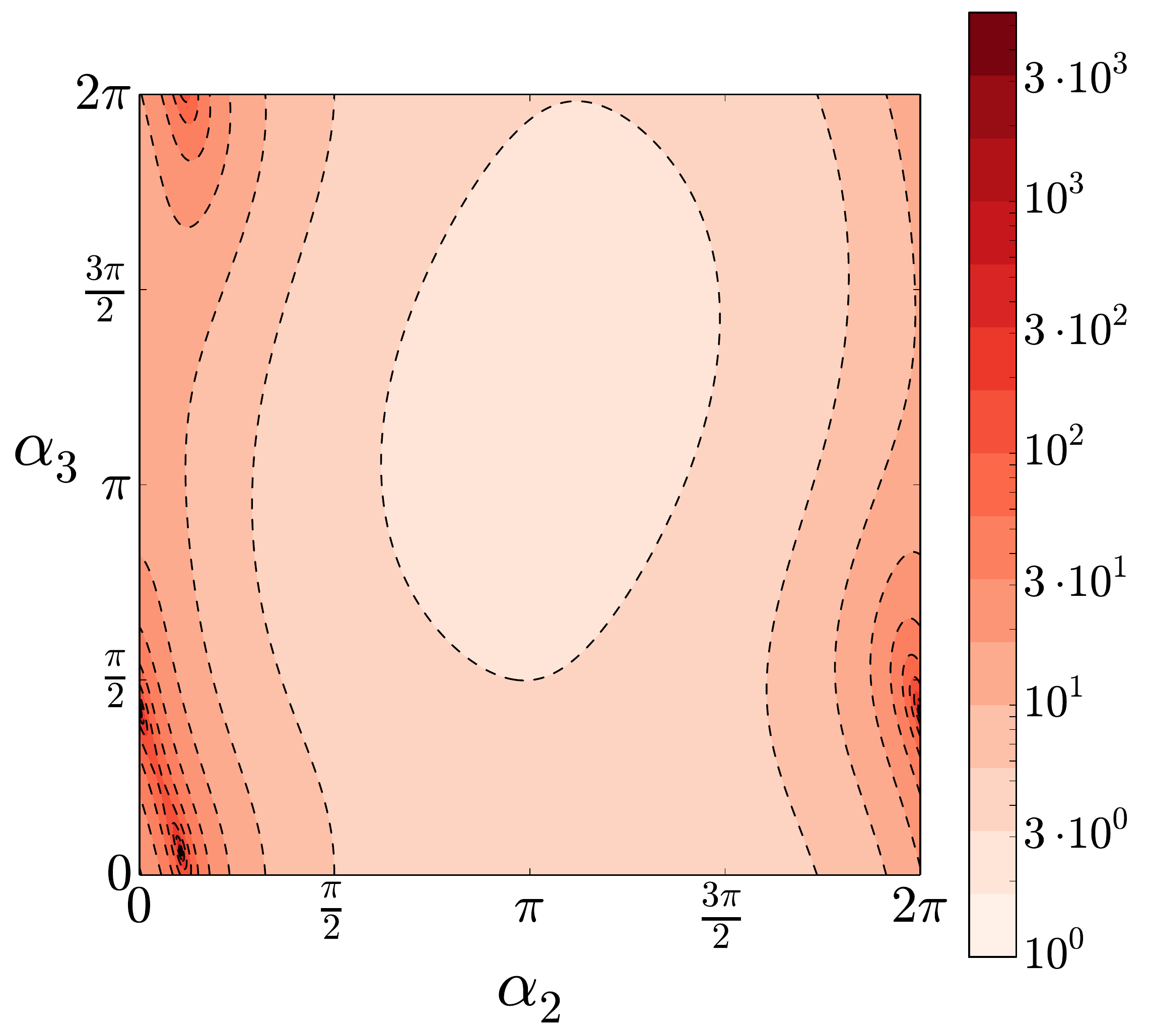}
	\caption{Upper bounds on the NSI coefficients $\varepsilon_{ee}$ (left), $\varepsilon_{e\mu}$ (centre) and $\varepsilon_{e\tau}$ (right) all multiplied by $10^{9}$ as a function of the Majorana phases $\alpha_2$ and $\alpha_3$, found from the $^{76} \mathrm{Ge}$ $0\nu\beta\beta$ decay limit $T_{1/2}^{0\nu\beta\beta} >5.3\times 10^{25} $ y. Best fit values for the mixing angles $\theta_{12}$, $\theta_{13}$ and $\theta_{23}$, squared mass splittings $\Delta m_{12}^2$ and $\Delta m_{23}^2$, Dirac CP phase $\delta$ are taken in the NO, with a lightest neutrino mass of $m_1 = 0 ~\mathrm{eV}$.}
\label{fig:esector0nubb}
\end{figure}

We now set $T^{0\nu\beta\beta}_{1/2} >$  5.3 $\times 10^{25}$ y derived from the $^{76}$Ge experiment GERDA-II \cite{Agostini:2017iyd,Agostini:2018tnm}. While KamLAND-Zen set a more stringent limit of $T^{0\nu\beta\beta}_{1/2} >$  1.07 $\times 10^{26}$ y with $^{136}$Xe \cite{KamLAND-Zen:2016pfg}, the exact number we use is not crucial for the following constraints and discussion. Setting each $\varepsilon_{e\lambda}$ to be non-zero at a time, Eq. (\ref{eq:onubbexpand}) can be solved to find an upper bound on the coefficient as a function of $\alpha_2$ and $\alpha_3$. These are displayed in the contour plots of Fig. \ref{fig:esector0nubb} for a smallest neutrino mass of $m_1 = 0 $ eV in the NO scheme. The associated \textquoteleft best' and \textquoteleft worst' upper bounds are shown in Table \ref{table:iii}. It can be seen that regardless of the value of the Majorana phases, the upper bound on $|\varepsilon_{ee}|$, as found previously in the literature, is of order $10^{-9}$. For very finely tuned values of $\alpha_2$ and $\alpha_3$, however, $|\varepsilon_{e\mu}|$ and $|\varepsilon_{e\tau}|$ are unbounded. Comparing all upper bounds in the $e$ sector, we see that $0\nu\beta\beta$ decay is unequivocally the best method to probe $\varepsilon_{ee}$. For a large portion of the $(\alpha_2,\alpha_3)$ parameter space it is also better at constraining $\varepsilon_{e\mu}$ and $\varepsilon_{e\tau}$. However, for certain fine-tuned values of the phases these coefficients become unbounded and KamLAND can provide a better upper bound.

While $0\nu\beta\beta$ decay is certainly the most sensitive process to test for LNV as outlined above, it can only probe the $e$ sector, whereas other observables may shed light on other flavour coefficients. An interesting process in this regard is the LNFV conversion of captured muons in nuclei, $\mu^- + (Z,A) \to e^+ + (Z-2,A)$. Proposed many years ago by Pontecorvo \cite{Pontecorvo:1967fh}, it has gained recent interest due to the upcoming searches for the LNC but LFV muon conversion $\mu^- + (Z,A) \to e^- + (Z,A)$ by the COMET \cite{Adamov:2018vin} and Mu2e \cite{Bartoszek:2014mya} collaborations, which promise an increased experimental sensitivity by four orders of magnitude. While it is doubtful that the current limit $R^\text{Ti}_{\mu e} \lesssim 10^{-11}$ \cite{Kaulard:1998rb}
on the LNFV mode conversion rate can be improved in a similar fashion due to different background considerations \cite{Yeo:2017fej}, $\mu^- - e^+$ conversion is an important complementary probe to $0\nu\beta\beta$ decay.

To estimate the sensitivity of the LNFV $\mu^- - e^+$ conversion process on the LNV NSI coefficients considered in this paper, we follow the estimate in \cite{Geng:2016auy}. In this approach, and using our notation, the conversion rate is approximated as 
\begin{equation} \label{eq:mue-conversion}
R_{\mu e} \approx |\xi_{\mu e}|^2~ \frac{G_F^2}{2}\frac{Q^6}{q^2}~,
\end{equation}
where the effective parameter $\xi_{\mu e}$ is defined by
\begin{equation} \label{eq:mue-conversion2}
\begin{split}
\begin{aligned}
|\xi_{\mu e}|^2&\equiv \bigg|\sum_i \big(U^*_{e i}\gamma_{\mu i}+U^*_{\mu i}\gamma_{e i}\big)\bigg|^2 \\
&=\sum\limits_{\lambda}\Big(F^{(e)}_{\lambda}(\boldsymbol{\zeta},\alpha_2,\alpha_3)~\varepsilon^2_{\mu\lambda}+F^{(\mu)}_{\lambda}(\boldsymbol{\zeta},\alpha_2,\alpha_3)~\varepsilon^2_{e\lambda}\Big)\\
&~~~~+\sum\limits_{\lambda<\lambda'}\Big(G^{(e)}_{\lambda\lambda'}(\boldsymbol{\zeta},\alpha_2,\alpha_3)~\varepsilon_{\mu\lambda}\varepsilon_{\mu\lambda'}+G^{(\mu)}_{\lambda\lambda'}(\boldsymbol{\zeta},\alpha_2,\alpha_3) ~\varepsilon_{e\lambda}\varepsilon_{e\lambda'}\Big)~.
\end{aligned}
\end{split}
\end{equation}
The two terms in the first line of Eq. (\ref{eq:mue-conversion2}) take into account that the LNV NSI can be at the interaction vertex of either the incoming $\mu^{-}$ or the outgoing $e^{+}$ (the latter being shown in Fig. \ref{fig:LNVprocesses}). The only difference between the two diagrams is the exchange $U^*_{ei}\gamma_{\mu i}\leftrightarrow U^*_{\mu i}\gamma_{e i}$.

In Eq.~(\ref{eq:mue-conversion}), $q$ denotes the momentum scale of the intermediate neutrino in the process, $q \approx 100$~MeV, and $Q$ is the energy release of the emitted positron determining the size of the phase space with $Q \approx 15.6$~MeV \cite{Berryman:2016slh}. Due to the process being incoherent and partially going to excited final nuclear states, $Q$ also approximately convolutes the nuclear matrix element of this transition. Here, it should be emphasized that the relevant nuclear matrix elements have not been calculated in detail and Eq. (\ref{eq:mue-conversion}) can only be regarded as a very rough estimate of the order of magnitude of the conversion rate. Nevertheless, using the experimental limit Eq. (\ref{eq:mue-conversion}), we estimate a limit on $|\xi_{\mu e}|$ of order $|\xi_{\mu e}| \lesssim 10^4$. This is barely stringent enough for the EFT assumptions to be consistent, with the limit corresponding to an effective operator scale $\Lambda = (|\xi_{\mu e}| G_F)^{-1/2} \approx 3~\text{GeV} > q$. Even with the most optimistic improvement of the future sensitivity to $R_{\mu e} \lesssim 10^{-16}$ \cite{Berryman:2016slh}, the coefficient $|\xi_{\mu e}|$ will only be probed at $|\xi_{\mu e}| \sim 30$, corresponding to a scale $\Lambda_{\mathrm{NP}} \sim 50$~GeV.

We summarise the constraints on the individual coefficients in the $\mu$ sector (when each is taken to be non-zero) in Table \ref{table:iii}. These are also of order $10^{4}$. Similar to $0\nu\beta\beta$ decay, the coefficients with flavour indices not corresponding to the flavour of external charged leptons are unbounded for particular values of the Majorana phases. Because the functions preceding $\varepsilon^2_{\mu\lambda}$ in Eq. (\ref{eq:mue-conversion2}) are $F^{(e)}_{\lambda}(\boldsymbol{\zeta},\alpha_2,\alpha_3)$, the dependences on the Majorana phases for the $\mu^{-} - e^{+}$  upper bounds are identical to those for $0\nu\beta\beta$ decay in Fig. \ref{fig:esector0nubb}, except being scaled by a factor of $\sim 10^{12}$. From Eq. (\ref{eq:mue-conversion2}) it is clear that $\mu^{-} - e^{+}$ can also probe the $e$ sector coefficients, but sets bounds larger than $0\nu\beta\beta$ decay by a factor of $10^{12}$. However, it is of small interest to note that the dependence on the Majorana phases is the same as that for the MINOS upper bounds because the functions preceding $\varepsilon^2_{e\lambda}$ in Eq. (\ref{eq:mue-conversion2}) are $F^{(\mu)}_{\lambda}(\boldsymbol{\zeta},\alpha_2,\alpha_3)$.

From the rare LNV meson decays such as $K^{\pm}\rightarrow\pi^{\mp}\mu^{\pm}\mu^{\pm}$ and  $B^{+}\rightarrow D^{-}\mu^{+}\mu^{+}$ and rare $\tau$ decays such as $\tau^{-}\rightarrow \pi^{-}\pi^{-}\mu^{+}$ it is also possible to probe the LNV coefficients considered in this work as well as those in the $\tau$ sector \cite{Dib:2000ce,Delepine:2009qg,Geib:2016atx}. However, at present these processes have sensitivities at or worse than that of $\mu^{-}-e^{+}$ conversion. Comparing the constraints in the $\mu$ sector in Table \ref{table:iii}, we see that the constraints from MINOS (and similar future LBL oscillations experiments) are currently far more stringent than microscopic LNV processes.

To conclude this section we will briefly mention the analyses in Refs. \cite{Geng:2016auy,deGouvea:2007qla} on the contribution toward the Majorana neutrino mass $M_{\alpha\beta}$ from the operators in Table \ref{table:i}. If the discussion is shifted back to the Wilson coefficients of the operators, i.e. we set
\begin{equation}
\varepsilon_{\beta\alpha} = \frac{v}{G_F}\frac{C_{\beta\alpha}}{\Lambda_{\mathrm{NP}}^3}~,
\end{equation}
and assuming that the Wilson coefficients $C_{\beta\alpha}\sim\mathcal{O}(1)$, the minimum sum of neutrino masses $\sum m_\nu > 6\times 10^{-2}$ eV sets a lower bound on $\Lambda_{\mathrm{NP}}$. For the scalar and tensor Dirac structures the loop mass contains one quark Yukawa coupling, and $\Lambda_{\mathrm{NP}} > 1\times 10^{6}$ TeV. For a vector Dirac structure the loop gets a contribution from the charged lepton Yukawa coupling and both quark Yukawa couplings if they are right-handed, giving $\Lambda_{\mathrm{NP}} > 6\times 10^{3}$ TeV. This can be decreased significantly to $\Lambda_{\mathrm{NP}} \sim 1$ TeV if the couplings to third generation quarks are suppressed by some flavour symmetry \cite{Geng:2016auy}.

\vspace{-0.5em}
\section{CONCLUSIONS}
\vspace{-0.5em}

In this paper, we have investigated the effect of lepton number violating non-standard interactions on long-baseline neutrino oscillations. If the light active neutrinos are of Majorana nature the \textquoteleft$\nu_\alpha\rightarrow\bar{\nu}_\beta$' process become possible, either through a $ (m_{\nu}/E_{\nu})^{2} $ suppressed light neutrino helicity reversal or an LNV charged current (CC)-like interaction at production or detection. The Majorana neutrino case is favoured from a model-building perspective and is currently being probed by a wide range of neutrinoless double beta ($0\nu\beta\beta$) decay experiments.

We first studied the different derivations of neutrino oscillations in quantum mechanics and quantum field theory. The QFT model is a more complete and physically consistent picture, taking into account the coherence of overlapping wave packets at production and detection.  In this picture we derived the $ (m_{\nu}/E_{\nu})^{2} $ suppression associated with the Majorana helicity reversal, showing that the total rate in this case cannot be factorised into a production flux, oscillation probability and detection cross section. In the QFT picture we next studied the LNV interaction at the detection process. It is immediately clear that if the chirality of the production and detection processes are opposite, an \textquoteleft oscillation probability' $ P_{\nu_{\alpha}\rightarrow\bar\nu_{\beta}} $ can be factorised from the total rate of the process, justifying the use of a simple non-standard oscillation formula. The $ (m_{\nu}/E_{\nu})^{2} $ suppression is replaced by a $ |\varepsilon|^{2} $ suppression, where $ \varepsilon $ parametrises the strength of the LNV NSI compared to the Fermi coupling $ G_{F} $.

Using a bound made by the MINOS experiment on \textquoteleft$\nu_{\mu}\rightarrow\bar{\nu}_\mu$' oscillations, we put limits on the $ \varepsilon $ flavour coefficients in the case of a LH SM interaction at production and a RH leptonic current (connected in turn to a LH or RH hadronic current) at detection. The limits are also valid for the LNV NSI being at production. In the two-neutrino mixing scheme which is approximately valid for the $\nu_{\mu}-\nu_{\tau}$ sector, we derived a simplified expression for the non-standard oscillation probability in terms of the coefficients $ \varepsilon_{\mu\mu} $ and $ \varepsilon_{\mu\tau} $. Multiplying the non-standard oscillation probability by the NuMI beam flux and quasi-elastic CC cross section, and integrating over the energy range $0-20$ GeV, we derived upper bounds on the absolute values of the two coefficients. While the value of the single Majorana CPV phase $ \eta $ alters the upper bounds somewhat, we conservatively obtained $ |\varepsilon_{\mu\mu}|\lesssim 0.6 $ and $ |\varepsilon_{\mu\tau}|\lesssim 0.7 $. If a future experiment like MINOS were to be at a smaller baseline we found that the value of $ \eta $ has a larger impact on the upper bound.

We subsequently generalised the constraints to the three-neutrino scheme, using the best fit values for the 3$\nu$ mixing parameters (including the most recent hinted value of the Dirac phase $\delta$) and exploited the MINOS bound to constrain the $\mu$ sector parameter space, $(\varepsilon_{\mu e}, \varepsilon_{\mu\mu}, \varepsilon_{\mu\tau}) $. Likewise, we used a KamLAND measurement limiting the number of solar $\bar{\nu}_e$ from the source of solar $^8$B $\nu_e$ to place constraints on the $e$ sector parameter space, $(\varepsilon_{e e}, \varepsilon_{e \mu},\varepsilon_{e\tau}) $. For the latter we took into account the MSW effect and decoherence of the propagating neutrinos. We again found upper limits on the absolute values of the flavour coefficients but which do not vary as a function of the Majorana phases $\alpha_2$ and $\alpha_3$.  We briefly mentioned that a future high statistics experiment above the $\tau$ threshold (and with a magnetic field in the far detector like OPERA) could put constraints on the  $\tau$ sector parameter space, $(\varepsilon_{\tau e},\varepsilon_{\tau \mu},\varepsilon_{\tau\tau})$.

Of course, as has been considered before in the literature, these coefficients can be constrained by experimental searches of microscopic LNV processes. In order to compare with the KamLAND results we used the most recent $0\nu\beta\beta$ decay half-life from the $^{76}$Ge experiment GERDA-II to put upper bounds on the electron sector. While $0\nu\beta\beta$ decay still provides the most competitive constraint on $|\varepsilon_{ee}|$, for particular values of the Majorana phases $|\varepsilon_{e\mu}|$ and $|\varepsilon_{e\tau}|$ become unbounded, where otherwise KamLAND sets a finite upper bound. Similarly, $\mu^- - e^+$ conversion sets very loose bounds on the muon sector coefficients -- the upper bounds from MINOS are superior for all values of the Majorana phases. We have focused on the LNV signal observables, but in general the LNV NSI can be constrained along with LNC NSI in processes that are insensitive to lepton number, such as charged pion decay. Both LNC and LNV CC-like NSI can induce deviations from lepton universality and CKM unitarity.

While the CPV properties of \textquoteleft$\nu_{\alpha}\rightarrow\bar{\nu}_{\beta}$' oscillations were considered in Ref.~\cite{Xing:2013ty} in the context of the light neutrino helicity reversal mechanism, future work could study the CPV properties of complex $ \varepsilon $ coefficients. It would also be interesting to consider phenomenology of LNV NC-like matter oscillations, which also generically arise from dimension-six operators in the LEFT of the  SM. Constraints on extensions of the 3$\nu$ scheme to additional sterile states may also be possible given the improving precision and possible charge-sign sensitivity of future SBL reactor experiments.

\vspace{-0.5em}
\section*{ACKNOWLEDGEMENTS}
\vspace{-0.5em}

P. D. B.  and F. F. D. would like to acknowledge support from the Science and Technology Facilities Council (STFC).

\bibliography{References}

\end{document}